\documentclass[10pt,fleqn,a4paper]{article}

\usepackage{grffile,color}
\usepackage{subfigure}
\usepackage{graphicx}
\usepackage{latexsym}
\usepackage{amsfonts, amssymb, amsmath}
\newcommand{\qed}{{\hfill $\square$}}
\newcommand{\e}[1]{\,{\rm e}^{#1}\,}
\newcommand{\caS}{{\mathcal S}}
\newcommand{\bbP}{{\mathbb P}}
\newcommand{\bbZ}{{\mathbb Z}}
\newcommand{\bea}{\begin{eqnarray}}
\newcommand{\eea}{\end{eqnarray}}
\newcommand{\bef}{\begin{figure}}
\newcommand{\enf}{\end{figure}}
\newcommand{\ball}{\begin{array}{ll}}
\newcommand{\bacl}{\begin{array}{cl}}
\newcommand{\bal}{\begin{array}{l}}
\newcommand{\bac}{\begin{array}{c}}
\newcommand{\ea}{\end{array}}

\newcommand{\Z}{{\mathbb{Z}}}
\newcommand{\E}{{\mathbb{E}}}
\newcommand{\bbE}{{\mathbb{E}}}

\renewcommand{\P}{{\mathbb{P}}}
\renewcommand{\bbP}{{\mathbb{P}}}

\newtheorem{theorem}{Theorem}[section]

\newtheorem{proposition}[theorem]{Proposition}

\newtheorem{conjecture}{Conjecture}

\begin{document}

\title{Lattice permutations and Poisson-Dirichlet distribution of cycle lengths}
\author{Stefan Grosskinsky$^{1,2}$, Alexander A. Lovisolo$^1$, Daniel Ueltschi$^2$}
%Email: {\tt S.W.Grosskinsky@warwick.ac.uk}

%\sloppy
\maketitle

\begin{abstract}
\noindent
We study random spatial permutations on $\Z^3$ where each jump $x\mapsto\pi (x)$ is penalized by a factor $\e{-T\| x-\pi (x)\|^2}$. The system is known to exhibit a phase transition for low enough $T$ where macroscopic cycles appear.
We observe that the lengths of such cycles are distributed according to Poisson-Dirichlet. This can be explained heuristically using a stochastic coagulation-fragmentation process for long cycles, which is supported by numerical data.
\end{abstract}

%\tableofcontents %This should be removed later

\section{Introduction}
\footnotetext[1]{Centre for Complexity Science, University of Warwick, Coventry CV4 7AL, United Kingdom}
\footnotetext[2]{Department of Mathematics, University of Warwick, Coventry CV4 7AL, United Kingdom}

Random permutations are common in probability theory and combinatorics \cite{ABT}. They also occur in statistical mechanics, albeit with an additional spatial structure. With $\Lambda$ denoting a finite box in $\bbZ^{3}$, we consider the set $\caS_{\Lambda}$ of permutations of $\Lambda$, i.e., the set of bijections $\Lambda \to \Lambda$. The probability of a given permutation $\pi \in \caS_{\Lambda}$ depends on the jump lengths in such a way that all sites are mapped in their neighborhoods. In this paper, we study the model with probability
\bea\label{dist}
\P_\Lambda (\pi )=\frac{1}{Z_\Lambda} \exp\Big( -T\sum_{x\in\Lambda} \|x-\pi (x)\|^2\Big)\ .
\eea
Here $T > 0$ is a positive parameter and $\|x-\pi(x)\|$ denotes the Euclidean distance between $x$ and $\pi(x)$. The normalization $Z_\Lambda$ is defined by
\bea\label{pf}
Z_\Lambda =\sum_{\pi\in \caS_\Lambda} \exp\Big( -T \sum_{x\in\Lambda} \|x-\pi (x)\|^2\Big)\ .
\eea
This model has its origin in Feynman's approach to the quantum Bose gas \cite{Fey}, where $T$ is proportional to the temperature. Bosons are described by Brownian trajectories with $\frac{1}{T}$ playing the r\^ole of time; this suggests the weights \eqref{dist} for the permutation $\pi$. The presence of a lattice is not really justified, but it does not affect the qualitative behavior, at least in dimension 3 (we comment on dimension 2 at the end of the article).

Let us understand the qualitative behavior of the model when we vary the parameter $T$.
The most probable permutation is the identity, which has weight $1$. Typical permutations should be close to the identity when $T$ is large, with small cycles here and there.
The weight in \eqref{dist} penalizes large jumps and we expect that $\|x -
\pi(x) \| \lesssim \frac{1}{\sqrt{T}}$ . As $T$ decreases, sites are allowed to
be mapped to more locations and the lengths of permutation cycles grow. One
expects that a phase transition takes place (in dimension 3 or more) that is
accompanied by the occurrence of infinite cycles. See Fig. \ref{fig
spatial permutation} for a schematic spatial permutation with small $T$. 
\begin{figure}[htb]
\centering
\includegraphics[height=6cm]{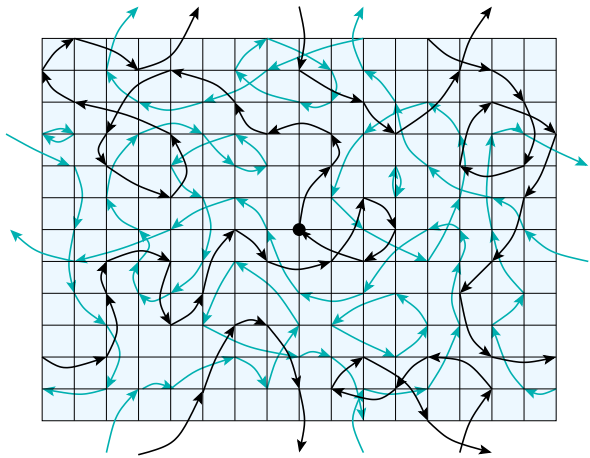}
\caption{\small A typical spatial permutation with periodic boundary conditions for small $T$. The cycle that contains the origin is depicted in black and it may be long. Isolated sites belong to 1-cycles (i.e., they are mapped onto themselves).}
\label{fig spatial permutation}
\end{figure}

The phase transition was observed numerically in \cite{GRU,Kerl}; it takes place at $T_{\rm c} \approx 1.71$. The fraction of sites that belong to macroscopic cycles was seen to converge to a non-random function $\nu_\infty (T)$ as $|\Lambda |\to\infty$, which is continuous and monotone decreasing in $T$. There are many macroscopic cycles and their sizes fluctuate. It was also observed that the average length of the largest cycle scales like $0.624 \nu_\infty (T) |\Lambda |$, which is identical to the expectation of the largest cycle in a random permutation with uniform distribution \cite{SL}. The latter observation was unexplained and puzzling at the time.

The situation is understood better now, and the explanation turns out to be surprisingly general. The joint distribution of the length of the long cycles is given by the {\it Poisson-Dirichlet distribution}.
This distribution has been introduced by Kingman in \cite{Kin1} and has cropped up in combinatorics, population genetics, number theory, Bayesian statistics and probability theory, see \cite{Feng,Kin2,ABT,PY,Pit} for details of applications and extensions.

Occurrence of the Poisson-Dirichlet distribution in models of statistical mechanics, i.e.\ in models with spatial structure, seems to have been noticed only recently. It has been rigorously established in the ``annealed'' model of spatial permutations where the locations of the sites are averaged upon \cite{BU2}. The proof is inspired by \cite{Suto1} and it uses a representation in terms of occupation numbers of Fourier modes, and non-spatial permutations within the modes. Such a structure is not present here, however.

We recall the definition of the Poisson-Dirichlet distribution in Section \ref{sec PD} and provide numerical evidence in Section \ref{sec num PD} that it is present in our model. In order to explain this, we show that the equilibrium state can be viewed as the stationary measure of an effective split-merge process. This strategy was recently applied successfully by Schramm to the random interchange model on the complete graph \cite{Sch} (the result was first conjectured by Aldous, see \cite{BD}). The absence of spatial structure makes the situation much simpler, but it was nonetheless a \textit{tour de force} to prove that long cycles occur, that they satisfy an effective  split-merge process, and that their asymptotic distribution is Poisson-Dirichlet (see also \cite{Bere} for subsequent simplifications and improvements). This strategy was also devised in \cite{GUW} for the cycles and loops that arise in the  T\'oth and Aizenman-Nachtergaele representations of quantum Heisenberg models in three spatial dimensions \cite{Toth,AN}. It allowed in particular to identify the parameter of the conjectured Poisson-Dirichlet distribution. Further situations that look similar include the random currents in the classical or quantum Ising models \cite{Aiz,CI,Gri}.

The key features are as follows: Long cycles are one-dimensional macroscopic objects and they are spread uniformly over the whole space. Introducing a suitable stochastic process with local changes, we observe that cycles are merged at a rate proportional to the number of ``contacts'' between them, and this number is proportional to the product of their lengths. Cycles are split at a rate proportional to the number of self-contacts, which is proportional to the square of the length. This is exactly analogous to a split-merge process on interval partitions \cite{Pit,Ald2,Bert}. As a consequence, the distribution of cycle lengths at equilibrium should be given by the invariant measure of the split-merge process, which is known to be the Poisson-Dirichlet distribution.

This explanation seems very attractive but it glosses over many technicalities. It assumes that a spatially uniform distribution of long cycles leads to a ``mean-field'' interaction and the correlations due to their spatial structure can be ignored. 
We provide mathematical background for these ideas in Section \ref{sec split-merge} and this allows us to state precise conjectures in Section \ref{sec eff split-merge}. These conjectures are confronted with numerical results in Section \ref{sec num split-merge}. As it turns out, the above heuristics is fully confirmed.

\section{Distribution of long cycles}

\subsection{Nature of long cycles}
\label{sec nature cycles}
Let us first give precise definitions for ``macroscopic'', ``mesoscopic'' and
``finite'' cycles in the infinite volume limit. Given $x \in \Lambda$ and a permutation $\pi\in \caS_\Lambda$, let $L_{x} (\pi )$ denote the length of the cycle that contains $x$, i.e., the number of sites in the support of this cycle.
\begin{itemize}
\item Macroscopic cycles occupy a non-zero fraction of the
volume. The fraction of sites in macroscopic cycles is given by
\begin{equation}
 \nu_{\rm macro}(T)=\lim_{\varepsilon \to 0^+} \liminf_{|\Lambda| \to \infty}
\frac{1}{|\Lambda|} \bbE_{\Lambda} \Bigl( \#\{ x \in \Lambda : L_x >\varepsilon|\Lambda|\} \Bigr).
\end{equation}
\item Mesoscopic cycles are infinite cycles that are not macroscopic.
The fraction of sites in mesoscopic cycles is given by
\begin{equation}
\nu_{\rm meso}(T) = \lim_{K \to \infty} \liminf_{|\Lambda| \to \infty}
\frac{1}{|\Lambda|} \bbE_{\Lambda} \Bigl( \#\{ x \in \Lambda : K < L_x < \frac{|\Lambda|}K \} \Bigr).
\end{equation}
\item Finally, the fraction of sites in finite cycles is given by
\begin{equation}
\begin{split}
\nu_{\rm finite}(T) &= \lim_{K \to \infty} \liminf_{|\Lambda| \to \infty}
\frac{1}{|\Lambda|} \bbE_{\Lambda} \Bigl( \#\{ x \in \Lambda : L_x < K \} \Bigr) \\
&= 1 - \nu_{\infty}(T).
\end{split}
\end{equation}
\end{itemize}
Here, $\nu_{\infty}(T) = \nu_{\rm meso}(T) + \nu_{\rm macro}(T)$ is the fraction of sites in infinite cycles.

\begin{figure}[htb]
        \begin{center}
        \subfigure[$T = 1.75 >T_c$]{%
           %\label{fig:second}
           \includegraphics[width=0.4\textwidth]{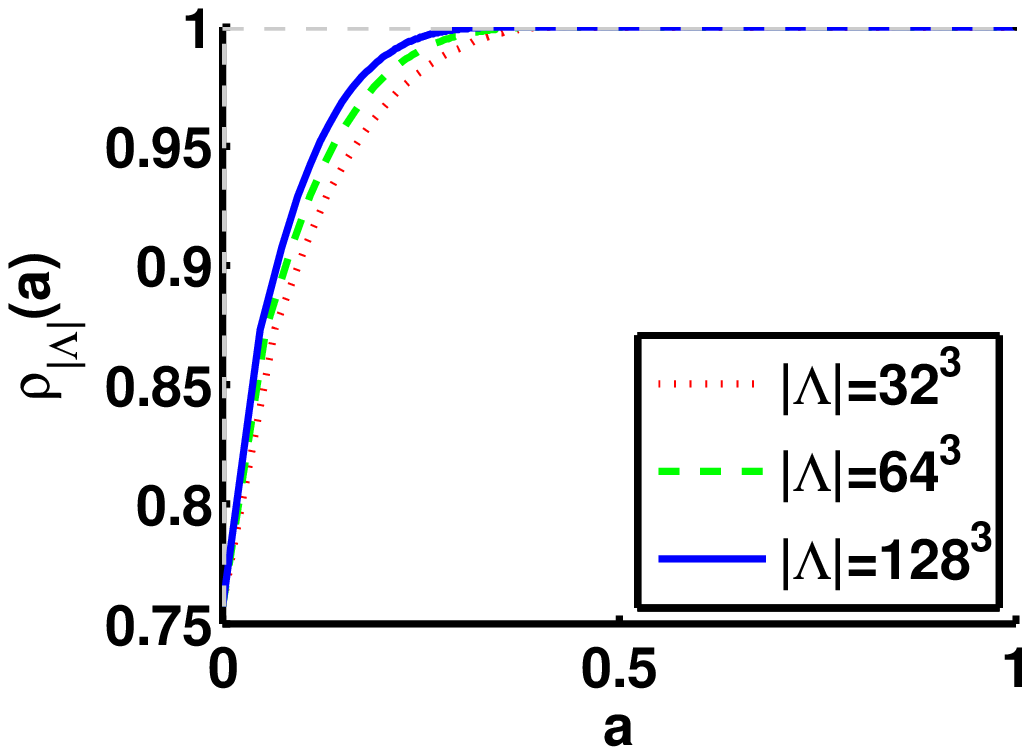}
        }
\qquad        \subfigure[$T=1.5 < T_c$]{%
            %\label{fig:first}
            \includegraphics[width=0.4\textwidth]{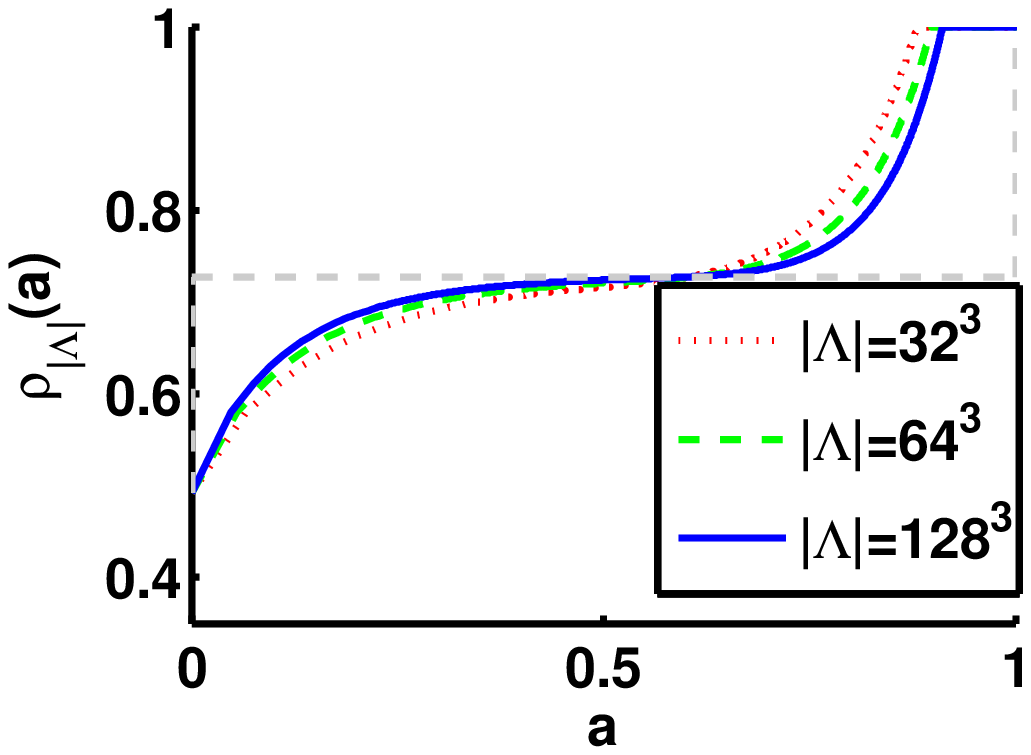}
        }\\
        \subfigure[$T=0.8 < T_c$]{%
            %\label{fig:first}
            \includegraphics[width=0.4\textwidth]{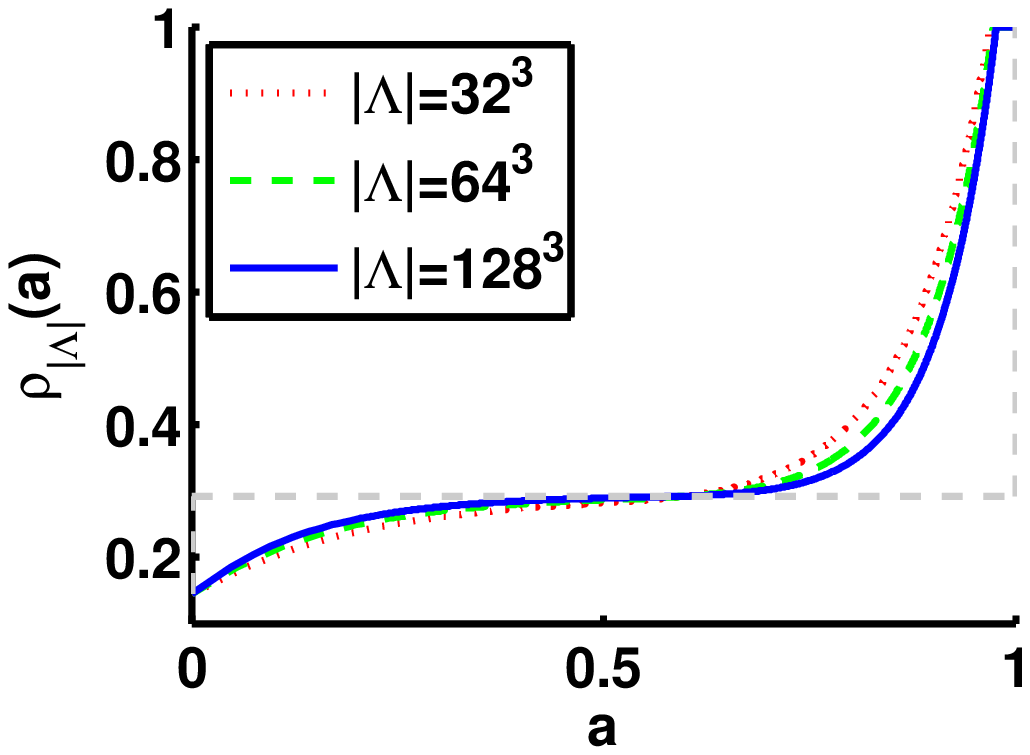}
        }
\qquad        \subfigure[$T=0.8$ (zoom)]{%
            %\label{fig:first}
            \includegraphics[width=0.4\textwidth]{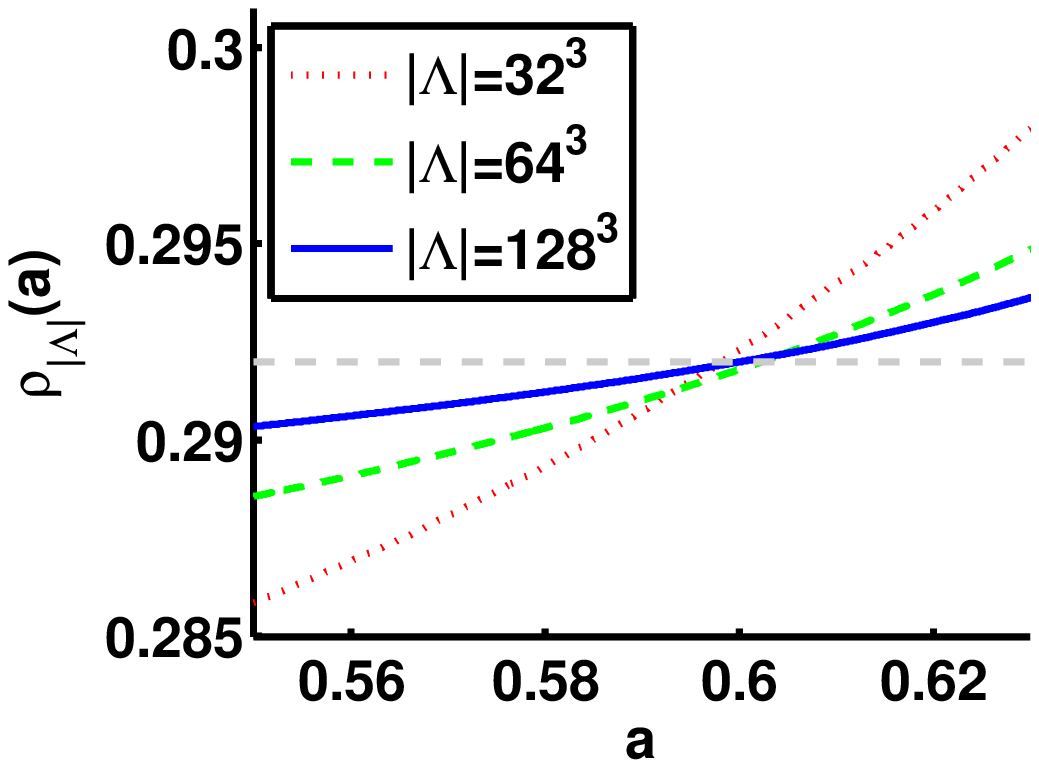}
        }
    \end{center}
    \caption{\small
    Plots of the expected fraction of sites $\rho_{|\Lambda|}(a)$ in cycles of
length
smaller than or equal to $|\Lambda|^{a}$. The horizontal dashed line indicates
$1-\nu_\infty(T)$, the fraction of particles in finite cycles in the
infinite volume limit. The curves have an intersection point independent of $|\Lambda|$ at $a \approx 0.6$ which is therefore used as the cutoff to distinguish long and short cycles. $\nu_\infty(0.8)$ is estimated to be $0.292$. Averages were taken over $5 \times 10^4$ realizations.
     }%
\label{fig:rho3D}
\end{figure}

One expects that only finite cycles are present when $T$ is large, that a phase with macroscopic cycles is present when $T$ is smaller than a positive number $T_{\rm c}$. %, and that no mesoscopic cycles occur. 
This was proved in the annealed model in \cite{BU1,BU2}. We check this numerically in the lattice model. Let
\begin{equation}
\rho_{|\Lambda |} (a) = \bbE_{\Lambda} \Bigl( \frac{\# \{ x \in \Lambda : L_{x} \leq |\Lambda|^{a} \}}{|\Lambda|} \Bigr).
\end{equation}
denote the fraction of sites that belong to cycles of length less than or equal to $|\Lambda|^{a}$. Notice that $\rho_{|\Lambda |} (0)$ is the fraction of particles mapped onto themselves and that $\rho_{|\Lambda |} (1) = 1$.
Numerical results for $\rho_{|\Lambda |} (a)$ are depicted in Fig.\ \ref{fig:rho3D}
for various parameters $T$.

In a finite domain, we need to define the cutoff that separates finite and
``infinite'' cycles. We can choose $|\Lambda|^{a}$ with \emph{any} power
$0<a<1$, since in the infinite volume limit, $\rho_{\infty}(a)$ will not depend
on our choice. The graphs of $\rho_{\Lambda}(a)$ depend on the size of the domain, but we see in Fig.\ \ref{fig:rho3D} (d) that they cross the same point around $a \approx 0.6$ for $T<T_c$. This value of $a$ is approximately independent of $T$ and we choose it for the cutoff, since it significantly reduces finite size effects. For the numerical results, we define the fraction $\nu_{|\Lambda |}$ of sites that belong to infinite cycles by
\begin{equation}
\nu_{|\Lambda |} (T) = 1 - \rho_{|\Lambda |} (0.6).
\end{equation}
In accordance with results for the annealed model \cite{BU1,BU2} we expect that $\nu_{\rm meso}(T) =0$ and that
\begin{equation}
\lim_{|\Lambda|\to\infty} \nu_{|\Lambda |} (T) = \nu_\infty (T) =\nu_{\rm macro}(T).
\end{equation}
This is supported by the numerics in Fig.~\ref{fig:rho3D}.

\subsection{Griffiths-Engen-McCloskey and Poisson-Dirichlet distributions}
\label{sec PD}

For a given permutation $\pi\in \caS_\Lambda$ we call the cycle at $x$ with length $L_x (\pi )$ \textit{macroscopic} if $L_x (\pi ) >|\Lambda |^{0.6}$, as discussed in the previous section. Let $L^{(1)} (\pi ),L^{(2)} (\pi ),\ldots ,L^{(k)} (\pi )$ denote the cycle lengths in decreasing order, where $L^{(k)}$ is the smallest macroscopic cycle for the permutation $\pi$. If $\lambda^{(i)} = \frac{L^{(i)}}{|\Lambda|}$ is the fraction of the sites in the $i$th macroscopic cycle, we define
\begin{equation}
\nu (\pi ):= \lambda^{(1)} + ... + \lambda^{(k)}
\end{equation}
to be the fraction of sites in macroscopic cycles, and have $\E_\Lambda (\nu ) = \nu_{|\Lambda |} (T)$. The sequence $(\lambda^{(i)})$ forms a random partition of the (random) interval $[0,\nu (\pi)]$. We now introduce the relevant measures on such partitions, that will allow us to
describe the joint distribution of cycle lengths. 

The Poisson-Dirichlet distribution (PD) is a one-parameter family but we only need the distribution with parameter $1$, so we ignore the paramter altogether. It is best introduced with the help of the Griffiths-Engen-McClosckey distribution (GEM).
The latter is also called the ``stick-breaking'' distribution. One can generate a random sequence of positive numbers $(\lambda_{1},\lambda_{2},\dots)$ such that $\sum_{i} \lambda_{i} = \nu$ as follows:
\begin{itemize}
\item choose $\lambda_{1}$ uniformly in $[0,\nu]$;
\item choose $\lambda_{2}$ uniformly in $[0,\nu-\lambda_{1}]$;
\item choose $\lambda_{3}$ uniformly in $[0,\nu-\lambda_{1}-\lambda_{2}]$;
\item and so on..., always chopping a piece off the remaining portion of the ``stick''.
\end{itemize}
This is equivalent to choosing a sequence of i.i.d.\ random variables $(\alpha_{1},\alpha_{2},\dots)$ where each $\alpha_{i}$ is taken uniformly in $[0,1]$, and then to form the sequence
\[
(\alpha_{1}, (1-\alpha_{1}) \alpha_{2}, (1-\alpha_{1}) (1-\alpha_{2}) \alpha_{3}, \dots) \times \nu.
\]

Our goal is to recognize that a given sequence has the distribution GEM. One can invert the above construction, and form an i.i.d.\ sequence out of a GEM sequence. Namely, if $(\lambda_{1},\lambda_{2},\dots)$ is GEM on the interval $[0,\nu]$, the following sequence is i.i.d.\ with respect to the uniform distribution on $[0,1]$:
\begin{equation}
\label{uniformiid}
(\alpha_{1},\alpha_{2},\alpha_{3},\dots) = \Bigl( \frac{\lambda_{1}}\nu, \frac{\lambda_{2}}{\nu - \lambda_{1}}, \frac{\lambda_{3}}{\nu - \lambda_{1} - \lambda_{2}}, \dots \Bigr).
\end{equation}

PD is a distribution on ordered partitions, and a PD sequence can be obtained by 
rearranging a GEM sequence in decreasing order. On the other hand, given an
ordered PD sequence, a GEM sequence can be obtained as a size-biased permutation
of that sequence \cite{Pit}.

\subsection{Numerical observations of cycle lengths}
\label{sec num PD}

The GEM distribution is easier to handle than the PD distribution, and it contains more information. We thus introduce an order on cycles allowing us to establish an order for the cycle lengths. This can be done as follows. First, choose an order for the sites of $\Lambda$. Then order the cycles according to the smallest sites in their support; namely, given two cycles $\gamma =(x_1 ,\ldots ,x_{|\gamma |} )$ and
$\gamma' =(x'_1 ,\ldots ,x'_{|\gamma' |} )$, we say that $\gamma <\gamma'$ if and only if
$\min_{1\leq i\leq |\gamma |} x_i <\min_{1\leq i\leq |\gamma' |} x'_i$. We then
denote $L_1 ,\, L_2 ,\ldots$ the lengths of cycles larger than $|\Lambda|^{0.6}$ in this order, which is not to be confused with the notation $L_x$ for lengths of cycles rooted in $x\in\Lambda$.

\begin{figure}[htb]
\centering
\begin{picture}(0,0)%
\includegraphics{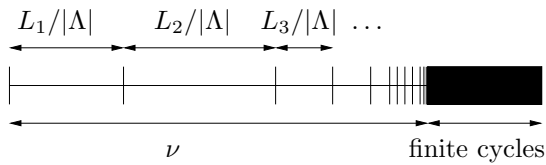}%
\end{picture}%
\setlength{\unitlength}{2072sp}%
\begingroup\makeatletter\ifx\SetFigFont\undefined%
\gdef\SetFigFont#1#2#3#4#5{%
  \reset@font\fontsize{#1}{#2pt}%
  \fontfamily{#3}\fontseries{#4}\fontshape{#5}%
  \selectfont}%
\fi\endgroup%
\begin{picture}(6413,1878)(2239,-2821)
\put(4096,-2716){\makebox(0,0)[lb]{\smash{{\SetFigFont{10}{12.0}{\rmdefault}{\mddefault}{\updefault}{\color[rgb]{0,0,0}$\nu$}%
}}}}
\put(6976,-2716){\makebox(0,0)[lb]{\smash{{\SetFigFont{10}{12.0}{\rmdefault}{\mddefault}{\updefault}{\color[rgb]{0,0,0}finite cycles}%
}}}}
\put(2341,-1186){\makebox(0,0)[lb]{\smash{{\SetFigFont{10}{12.0}{\rmdefault}{\mddefault}{\updefault}{\color[rgb]{0,0,0}$L_1/|\Lambda|$}%
}}}}
\put(3961,-1186){\makebox(0,0)[lb]{\smash{{\SetFigFont{10}{12.0}{\rmdefault}{\mddefault}{\updefault}{\color[rgb]{0,0,0}$L_2/|\Lambda|$}%
}}}}
\put(6301,-1186){\makebox(0,0)[lb]{\smash{{\SetFigFont{10}{12.0}{\rmdefault}{\mddefault}{\updefault}{\color[rgb]{0,0,0}$\ldots$}%
}}}}
\put(5221,-1186){\makebox(0,0)[lb]{\smash{{\SetFigFont{10}{12.0}{\rmdefault}{\mddefault}{\updefault}{\color[rgb]{0,0,0}$L_3/|\Lambda|$}%
}}}}
\end{picture}%
\caption[partitionNu]{\small The lengths of macroscopic cycles, divided by the volume, give a random partition of $[0,\nu]$ which is expected to follow the GEM distribution.}\label{fig:partitioning}
\end{figure}

Let $\nu \equiv \nu (\pi )$ for a given permutation. 
Our aim is to show that $(\frac{L_1}{\nu \left | \Lambda  \right
|}, \frac{L_2}{\nu \left | \Lambda \right |}...)$ converges to GEM as $|\Lambda |\to\infty$,
as illustrated in Fig.~\ref{fig:partitioning}. This is equivalent to showing that
\begin{equation}\label{alphadef}
(\alpha_{1},\alpha_{2},\alpha_{3},\dots) = \Bigl( \frac{L_1}{\nu \left |
\Lambda \right |}, \frac{L_2}{\nu \left | \Lambda 
\right | - L_1},  \frac{L_3}{\nu \left | \Lambda  \right | - L_1 - L_2}, ... \Bigr)
\end{equation}
converges to a sequence of i.i.d.\ uniform random variables in $[0,1]$, see Eq.\ \eqref{uniformiid}.

\begin{figure}[h!]
    
    \begin{center}
        \subfigure[$F_{\alpha_1}(s)$]{%
            %\label{fig:first}
            \includegraphics[width=0.4\textwidth]{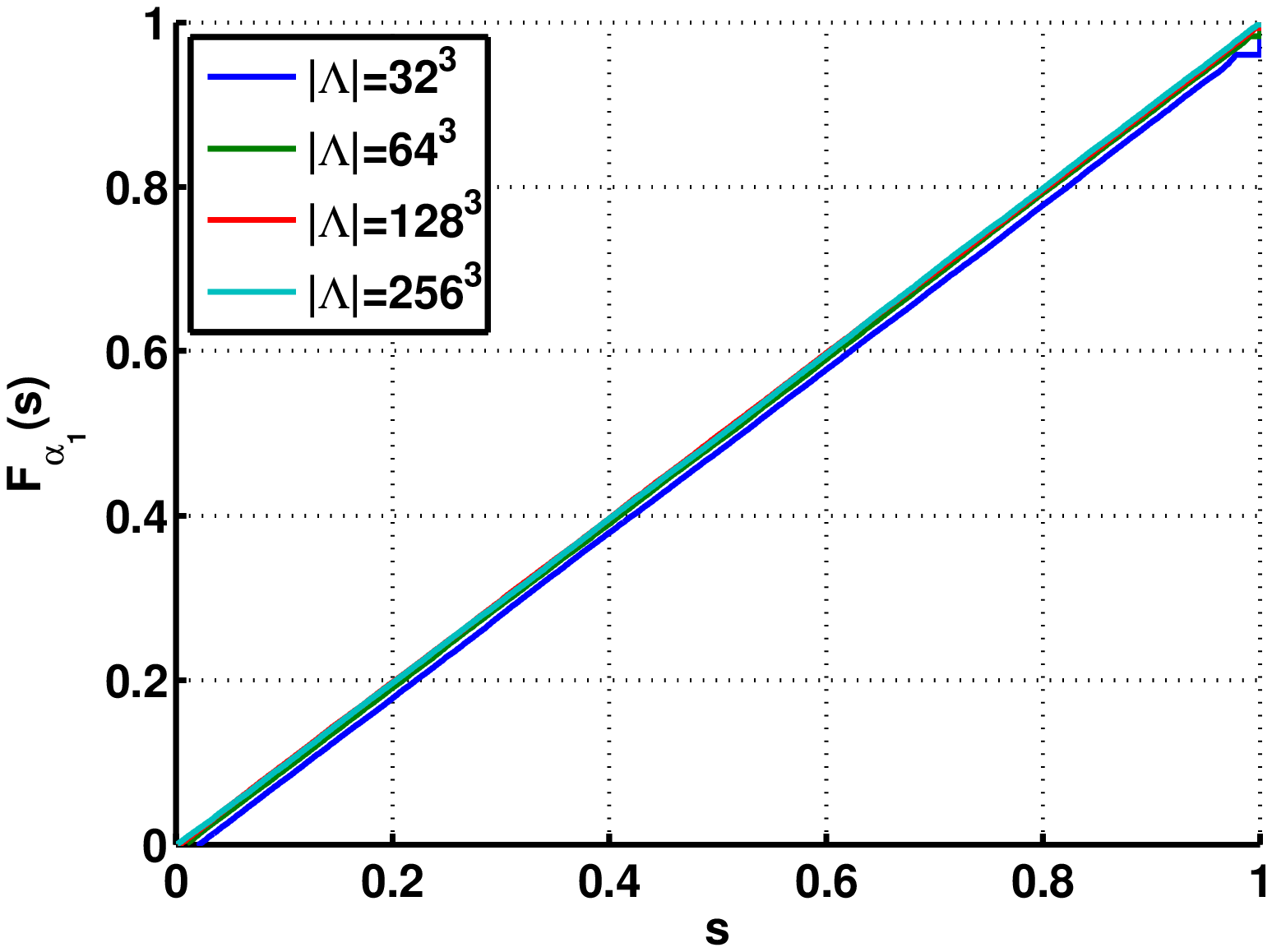}
        }%
\qquad        \subfigure[$F_{\alpha_1}(s)-s$]{%
           %\label{fig:second}
           \includegraphics[width=0.4\textwidth]{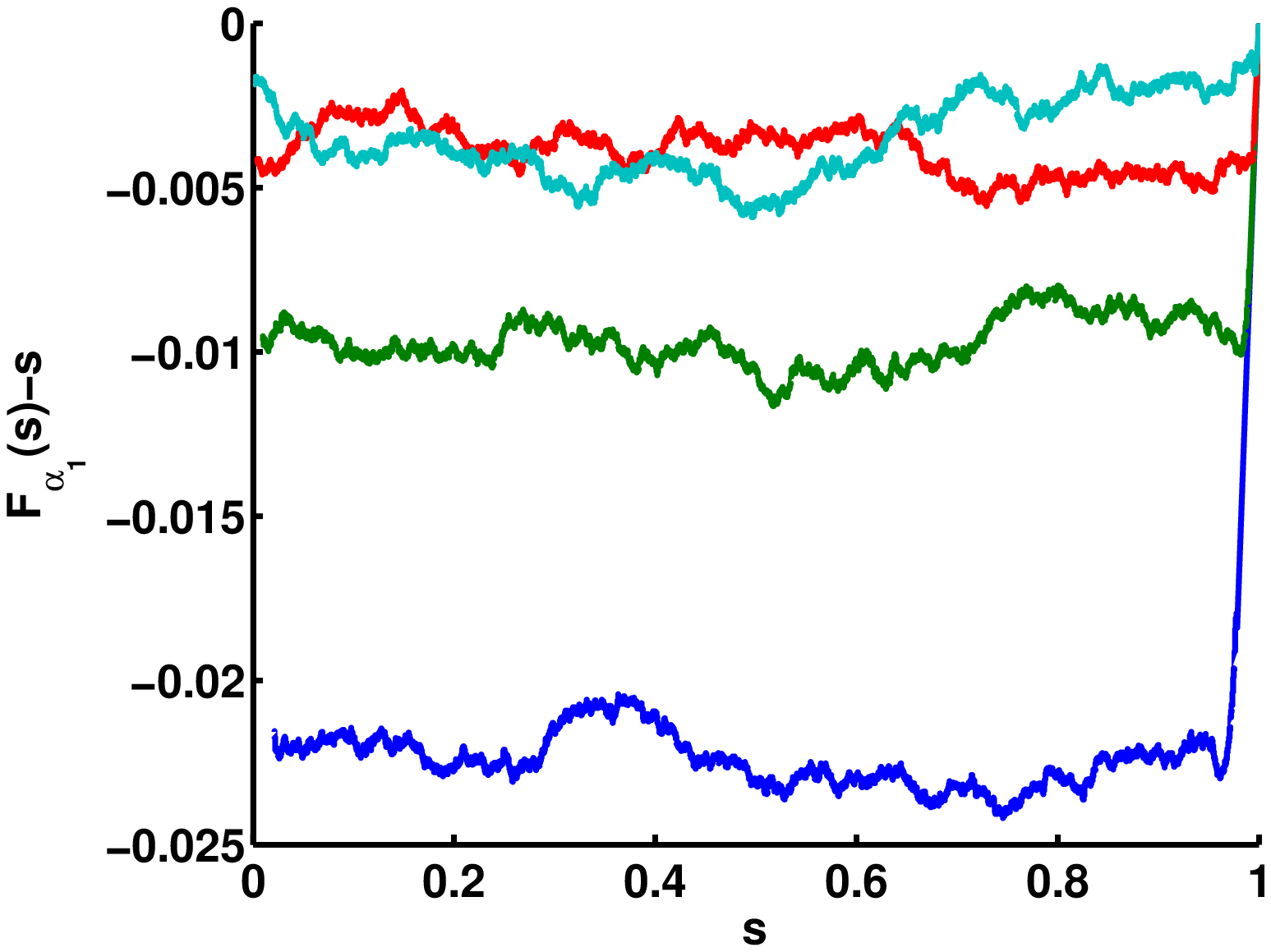}
        }\\ %  ------- End of the first row ----------------------%
\vspace{-2mm}
        \subfigure[$F_{\alpha_2}(s)$]{%
            %\label{fig:third}
            \includegraphics[width=0.4\textwidth]{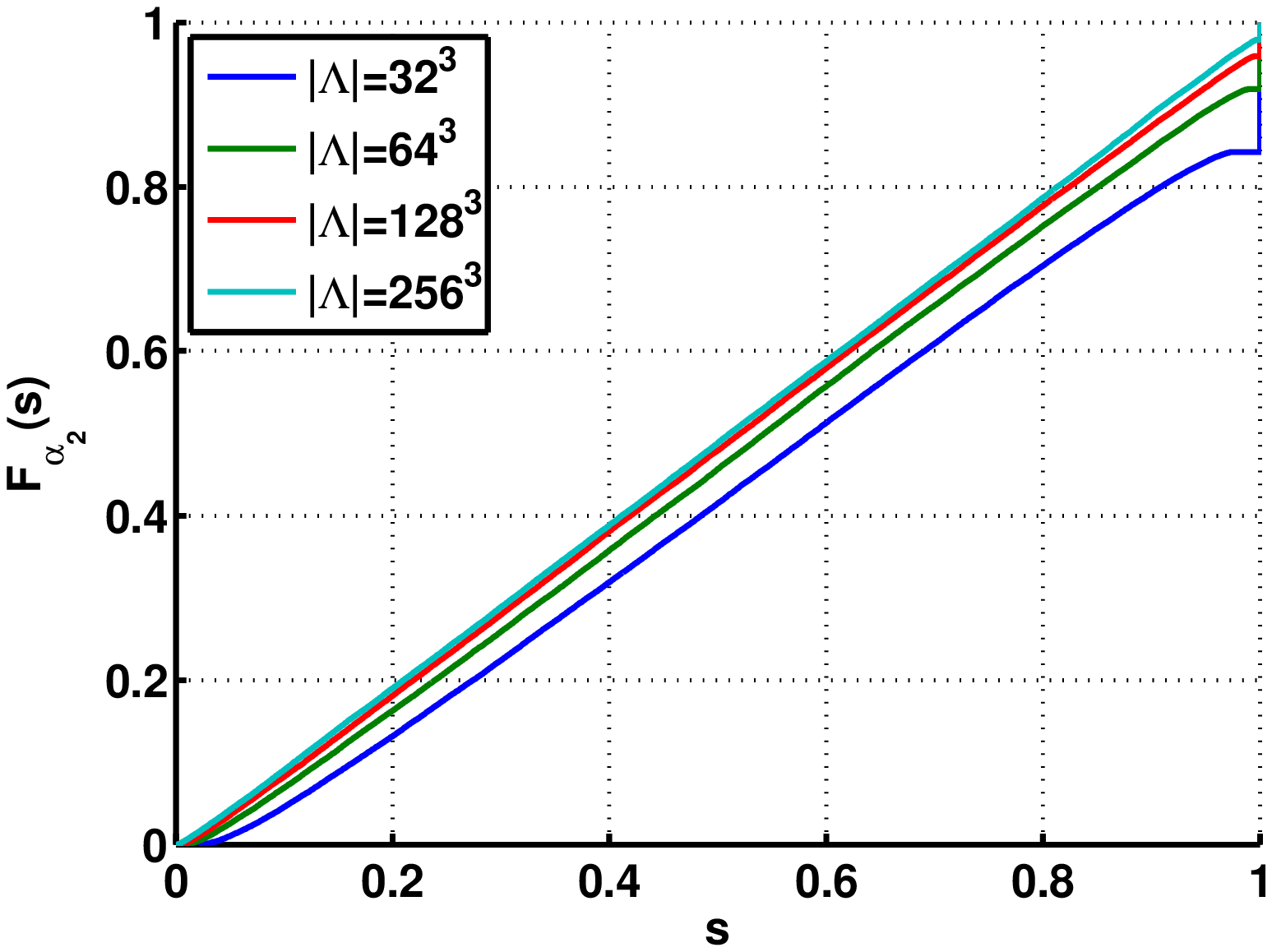}
        }%
\qquad        \subfigure[$F_{\alpha_2}(s)-s$]{%
            %\label{fig:fourth}
            \includegraphics[width=0.4\textwidth]{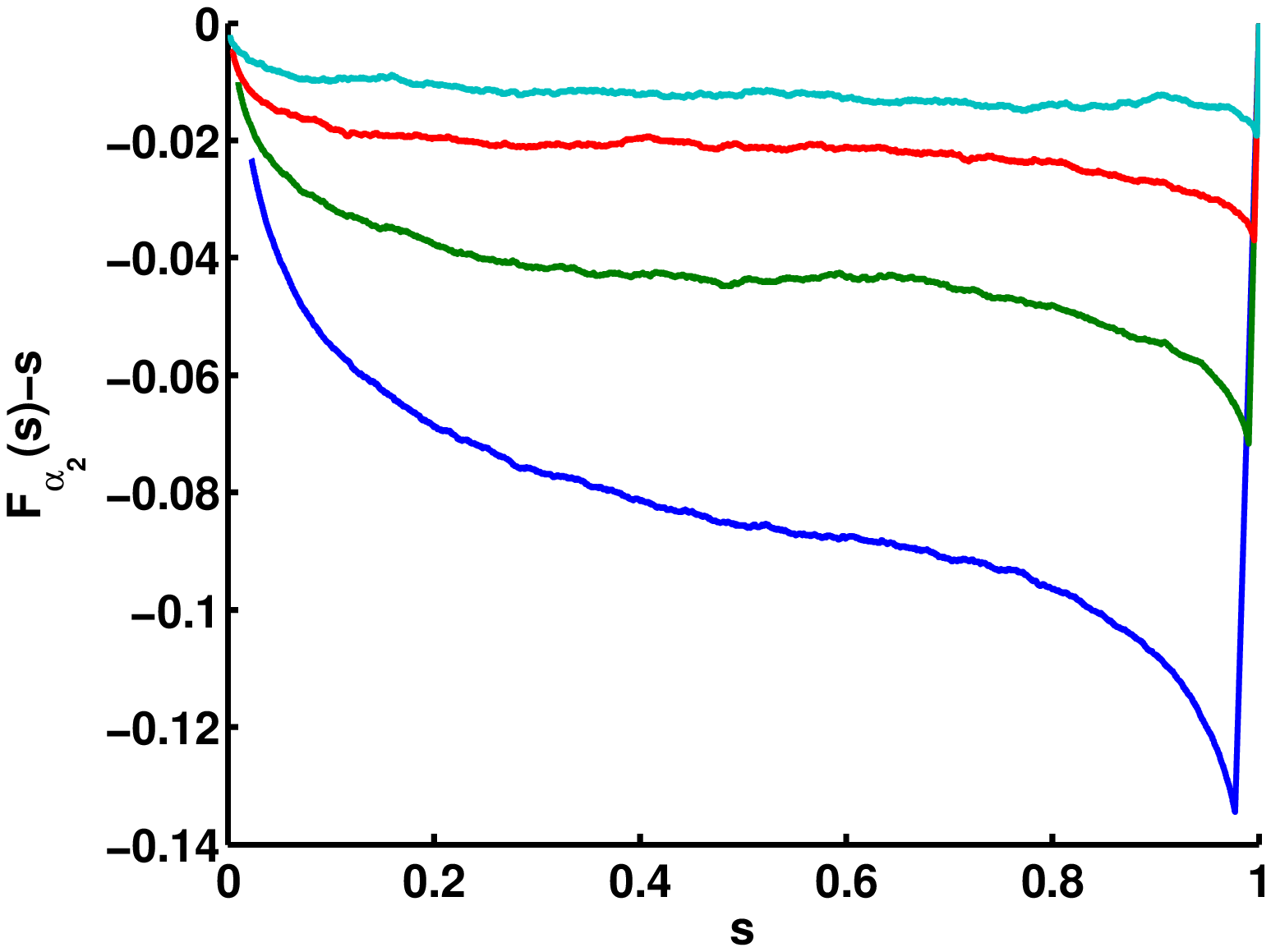}
        }\\%
\vspace{-2mm}
        \subfigure[$F_{\alpha_3}(s)$]{%
            %\label{fig:third}
            \includegraphics[width=0.4\textwidth]{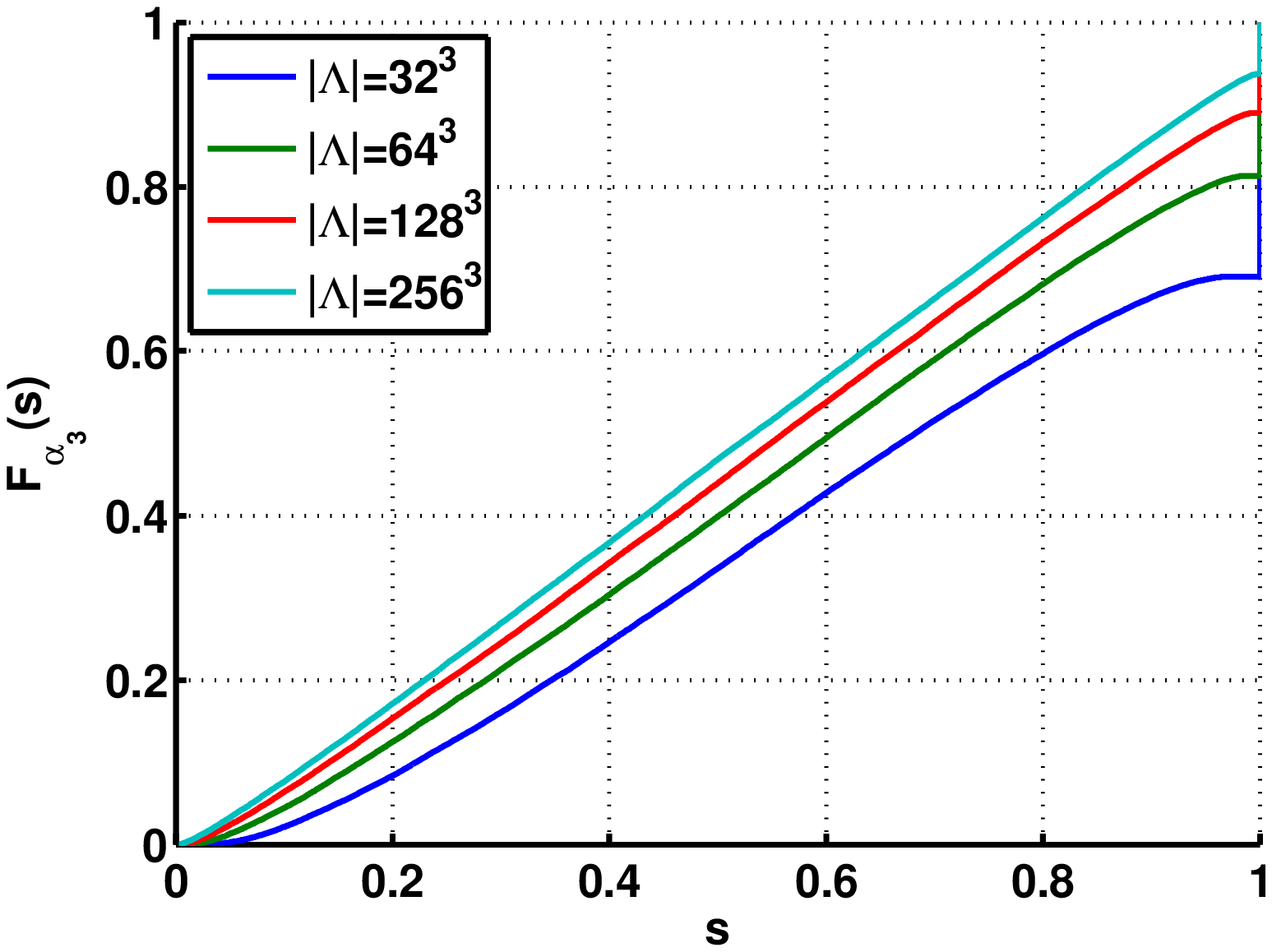}
        }%
\qquad        \subfigure[$F_{\alpha_3}(s)-s$]{%
            %\label{fig:fourth}
            \includegraphics[width=0.4\textwidth]{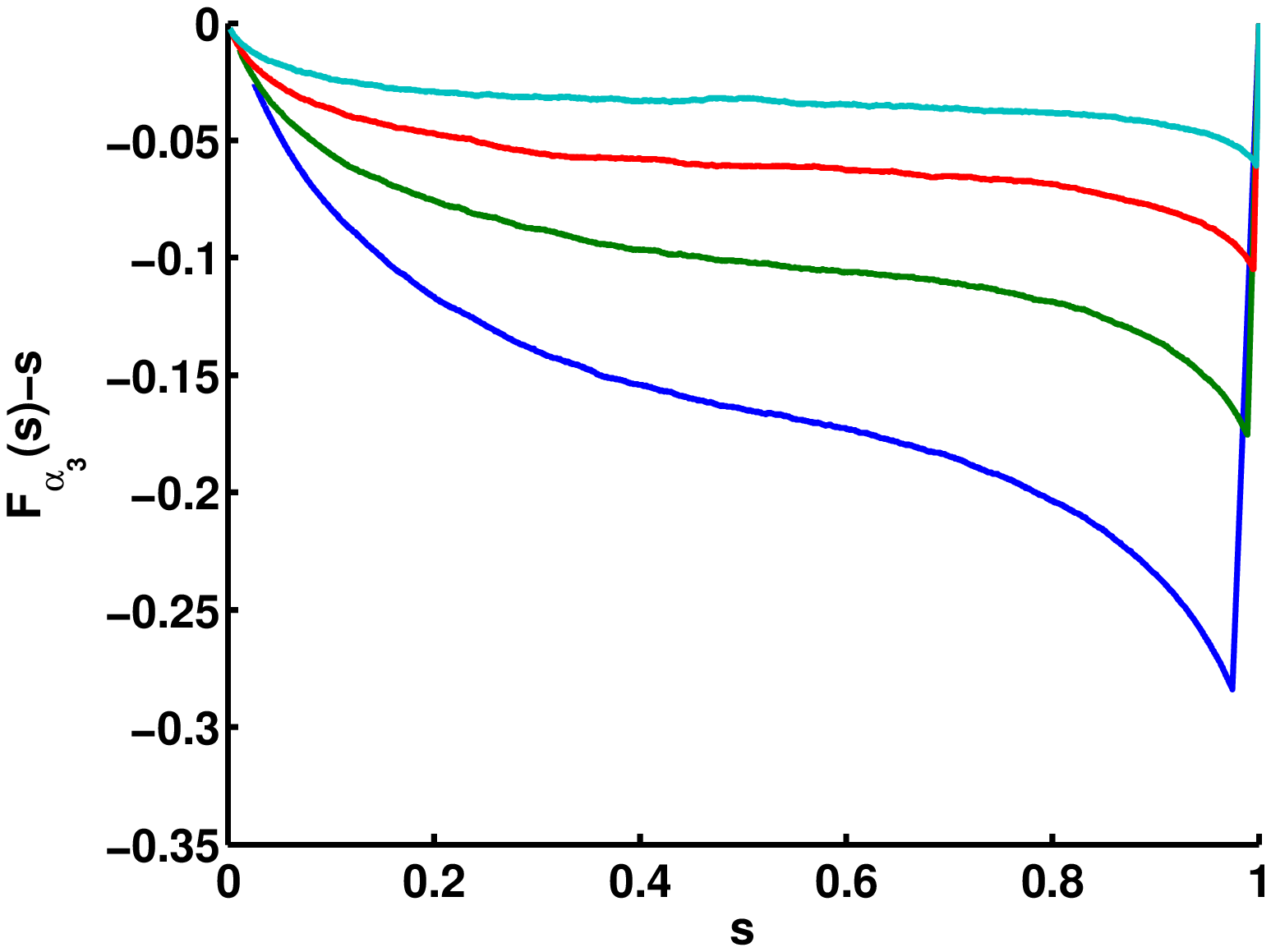}
        }%
    \end{center}
\vspace{-6mm}
    \caption{\small
        Cumulative distribution functions $F_{\alpha_i}(s)$ (\ref{alphadist}) with $i=1,2,3$
for system sizes $|\Lambda|=32^3,\ 64^3,\ 128^3,\ 256^3$ and $T=0.8$. As the volume tends to infinity, they converge to the CDF of a uniform random variable. Averages were taken over $10^5$ realizations.
     }%
\label{fig:cumulativegamma}
\end{figure}

\begin{figure}[htb]
\begin{center}
\includegraphics[width=0.6\textwidth]{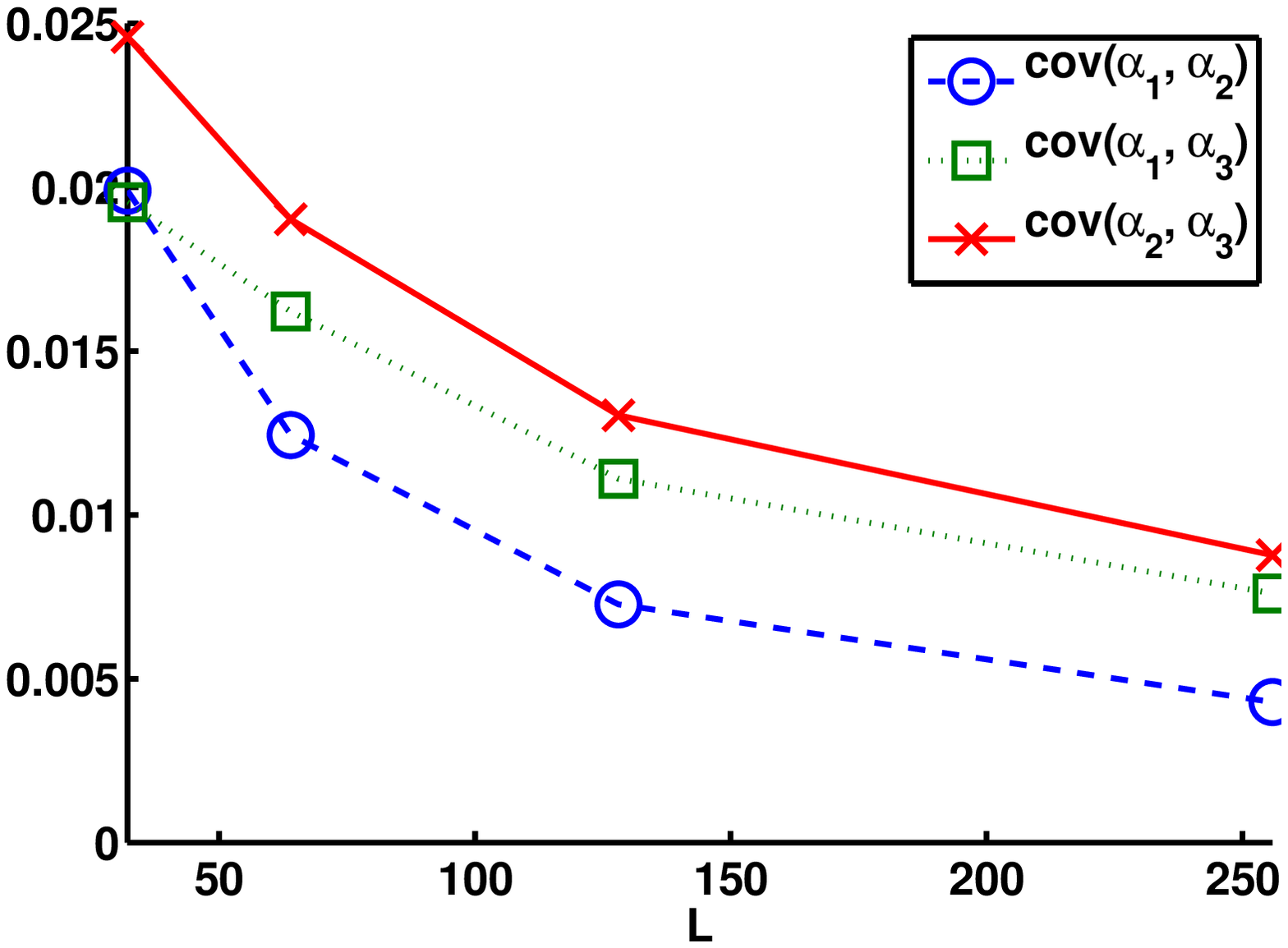}
 % means.png: 814x414 pixel, 90dpi, 22.97x11.69 cm, bb=0 0 651 331
\caption{\small Covariances of $\alpha_1, \alpha_2, \alpha_3$ as given in (\ref{alphadef}) shown for different system sizes $|\Lambda|=32^3,\ 64^3,\ 128^3,\ 256^3$ and $T=0.8$.}
\label{fig:covariances}
\end{center}
\end{figure}

\noindent The Cumulative Distribution Function for $\alpha_i$ is defined as
\bea\label{alphadist}
F_{\alpha_i}(s)= P(\alpha_i \leq s).
\eea
Numerical plots of $F_{\alpha_{i}}$ for the first three cycles can be found in Fig.~\ref{fig:cumulativegamma}. They clearly point to uniform random variables. Covariances can be found in 
Fig.~\ref{fig:covariances}, showing that they do indeed tend to $0$ in the infinite volume limit. The discontinuities at $s=1$ in Fig.~\ref{fig:cumulativegamma} for $i=1,2,3$ are due to the fact that in finite volumes it may happen that only $0,1,2$ cycles larger than $|\Lambda|^{0.6}$, respectively, are present.

\subsection{Markov chain Monte-Carlo}
\label{sec MCMC}

To sample spatial permutations we use a Markov chain Monte-Carlo process which
is ergodic and has $\bbP_{\Lambda}$ as its unique stationary distribution. Let
$B_{\Lambda}$ denote a suitable set of bonds, i.e., a set of unordered pairs
$\{x,y\}$ of sites $x,y \in \Lambda$. Let $\tau_{xy} =\tau_{yx}$ denote the
transposition of $x$ and $y$. We say that two permutations $\pi, \pi' \in
S_{\Lambda}$ are ``in contact'', noted $\pi \sim \pi'$, if there exists a bond
$\{x,y\} \in B_{\Lambda}$ such that $\pi' =\pi\circ\tau_{xy}$. Let $Q(\pi ,\pi'
)$, $\pi ,\pi'\in \caS_\Lambda$, be the transition matrix of a continuous-time Markov
chain $(\pi_t :t\geq 0)$ on $\caS_\Lambda$.

\begin{proposition}
Suppose that $B_\Lambda$ is large enough so that the graph $(\Lambda ,B_\Lambda )$ is connected, and that $Q(\pi ,\pi') > 0$ whenever $\pi \sim \pi'$. The Markov chain with transition rates $Q$ is ergodic.
\end{proposition}

\textsc{Proof.} This is done in \cite{Kerl}, but we recall the argument here. The space $\caS_\Lambda$ of lattice permutations is finite and irreducibility of the chain implies ergodicity. It then suffices to show that for all $\pi ,\pi'\in \caS_\Lambda$, $\pi'\neq\pi$ there exists $n<\infty$ such that $Q^n (\pi ,\pi' )>0$. Every permutation $\pi$ can be represented by a composition of transpositions. We still need to show that each of these transpositions can be written as a composition of transpositions along bonds of $B_{\Lambda}$.

Since the graph $(\Lambda ,B_\Lambda )$ is connected, there exists a connected path $(x_{0}, x_{1},\dots,x_{m})$ such that $x_{0}=x$, $x_{m}=y$, and $\{x_{i-1},x_{i}\} \in B_{\Lambda}$ for all $1 \leq i \leq m$. One can check that the following composition gives $\tau_{x,y}$:
% (draw a diagram!):
\begin{equation}
\label{transpositions}
\tau_{x,y} = \tau_{x_{0},x_{1}} \circ \dots \circ \tau_{x_{m-2},x_{m-1}} \circ \tau_{x_{m-1},x_{m}} \circ \tau_{x_{m-2},x_{m-1}} \circ \dots \circ \tau_{x_{0},x_{1}}.
\end{equation}
This shows that every $\pi\in \caS_\Lambda$ is connected to the identity permutation under the Markov chain dynamics.
\qed
\medskip

The composition of $\pi$ with a transposition is explicitly given by
\bea
\label{composition}
(\pi\circ\tau_{xy} )(z)=\pi\big(\tau_{xy} (z)\big) =\left\{\bacl \pi (z)\ &\mbox{if }z\neq x,y,\\ \pi (y)\ &\mbox{if }z=x,\\ \pi (x)\ &\mbox{if }z=y.\ea\right.\ 
\eea
Let $H_\Lambda (\pi )$ denote the ``energy'' of $\pi\in \caS_\Lambda$,
\bea\label{hamiltonian}
H_\Lambda (\pi )=T \sum_{x\in\Lambda} \|x-\pi (x)\|^2\ .
\eea
Here the Euclidean distance $\|\cdot\|$ is measured with periodic boundary conditions on a regular box $\Lambda\in\Z^3$. 
The distribution (\ref{dist}) then assumes the familiar form of the Gibbs state $e^{-H_\Lambda (\pi )}/Z_\Lambda$. For the transition rates we choose
\bea
\label{tp}
Q(\pi ,\pi' )=\begin{cases} \frac{1}{|B_\Lambda |} \min (1,\e{-(H_\Lambda (\pi')-H_\Lambda (\pi))}) &\text{if }\pi' \sim\pi, \\
0 &\text{otherwise.} \end{cases}
\eea
Note that all rates are in $[0,1]$ and can therefore be used as acceptance probabilities for the standard Metropolis algorithm: pick a bond $\{ x,y\}\in B_\Lambda$ uniformly at random and swap the images of $x$ and $y$ under $\pi$ with probability $1$ if this lowers the energy, and with probability $e^{-(H_\Lambda (\pi')-H_\Lambda (\pi))}<1$ if the swap increases the energy.

It is clear that the measure $\P_{\Lambda}$ fulfills the detailed balance conditions, since
\bea
e^{-H_\Lambda (\pi )} Q(\pi ,\pi')=e^{-H_\Lambda (\pi' )} Q(\pi' ,\pi)
\eea
for all $\pi ,\pi'\in \caS_\Lambda$. This implies stationarity.

The particular algorithm we use is the ``swap only'' method described in \cite{Kerl}.
The initial permutation is set to be the identity.
The Metropolis steps are then as follows: 
\begin{itemize}
\item Choose a bond $\{x,y\}$ of nearest-neighbors at random (we use periodic boundary conditions).
\item The candidate permutation, $\pi^\prime = \pi \circ \tau_{xy}$, replaces $\pi$ with probability $\min(1, e^{-(H_\Lambda (\pi')-H_\Lambda (\pi))})$.
\end{itemize}

The Metropolis step is then computationally fast since $H_\Lambda (\pi\circ\tau_{xy} )-H_\Lambda (\pi )$ depends only on local terms:
\bea\label{deltaH}
H_\Lambda (\pi\circ\tau_{xy} )-H_\Lambda (\pi )=T \Big(\| x{-}\pi (y)\|^2 +\| y{-}\pi (x)\|^2 -\| x{-}\pi (x)\|^2 -\| y{-}\pi (y)\|^2 \Big)\ .
\eea

\begin{figure}[htb]
\begin{center}
\includegraphics[width=0.6\textwidth]{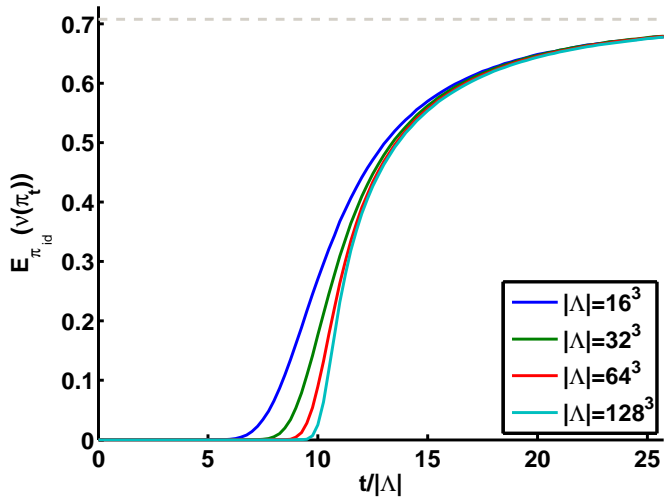}%
\end{center}
\caption{\small Expected value of the fraction of particles in macroscopic cycles
for the permutation after $t$ Metropolis steps starting from the identity
permutation, for $|\Lambda|=16^3,\ 32^3,\ 64^3,\ 128^3$ and $T=0.8$. Rescaling
time by
$\frac{1}{|\Lambda|}$ shows the asymptotic behavior, showing that the
equilibration time is of order $|\Lambda|$. The dashed line indicates the
asymptote for the curves, $\nu_\infty(0.8)$. Averages were taken over $10^4$ realizations.}
\label{fig mixingtime}
\end{figure}

We use a standard approach based on the ergodic theorem for sampling, where we
let the system equilibrate for a number of Metropolis steps of order $10^3|\Lambda |$, and
ensure that our measurements are spread over $5\times10^3 |\Lambda|$ steps.
We have strong numerical evidence that
the equilibration time of relevant observables is indeed of order $|\Lambda |$,
as is shown in Fig \ref{fig mixingtime}.

\subsection{Periodic boundary conditions}

We use periodic boundary conditions where $\Lambda\subset\Z^3$ is a 3-dimensional torus with equal side lengths $L$. This has the advantage of having less finite size effects than other choices such as closed boundary conditions. In the limit $|\Lambda|\to\infty$ we expect our results not to depend on that choice.

Precisely, for $y \in \bbZ^d$ we define $v_{i}(y)$ to be the $i$th component of $y$ modulo $L$, in such a way that $v_{i}(y) \in \{ -\frac L2 +1, \dots, \frac L2 \}$ (we assume here that $L$ is even, but the modifications for odd $L$ are straighforward). The Euclidean distance on $\bbZ^d$ is then replaced by
\begin{equation}
\|y\| = \Bigl( \sum_{i=1}^d |v_i(y)|^2 \Bigr)^{1/2}.
\end{equation}

Permutations on the torus can be charaterized by their {\it winding number}, which is in reality a winding vector. The winding number of $\pi$ in the $i$th direction, $i = 1,\dots,d$, is the integer
\begin{equation}
W_{i}(\pi) = \frac1L \sum_{x \in \Lambda} v_{i}(\pi(x)-x).
\end{equation}

In a large box, one should not expect any jumps of order $L$ for positive $T>0$ because of the Gaussian weights (\ref{dist}). The dynamics restricted to such permutations conserves the winding number:

\begin{proposition}
Suppose that the permutation $\pi$ satisfies
\[
\max_{x \in \Lambda} \| \pi(x) - x \| \leq \tfrac L2 - 2.
\]
Then $W_{i}(\pi \circ \tau_{xy}) = W_{i}(\pi)$ for all $i$ and all pairs $(x,y)$ of nearest-neighbors in $\Lambda$.
\end{proposition}

{\sc Proof.}
Using \eqref{composition}, we have
\begin{equation}
\label{diff winding}
W_{i}(\pi \circ \tau_{xy}) - W_{i}(\pi) = \tfrac1L v_{i}(\pi(y)-x) + \tfrac1L v_{i}(\pi(x)-y) - \tfrac1L v_{i}(\pi(x)-x) - \tfrac1L v_{i}(\pi(y)-y).
\end{equation}
Because of the modulo operation, we have
\begin{equation}
\begin{split}
&v_i(\pi(y)-x) = [\pi(y) - x]_{i} + k_1 L, \quad & v_i(\pi(x)-x) = [\pi(x) - x]_{i} + k_3 L, \\
&v_i(\pi(x)-y) = [\pi(x) - y]_{i} + k_2 L, & v_i(\pi(y)-y) = [\pi(y) - y]_{i} + k_4 L.
\end{split}
\end{equation}
Here, $[\cdot]_{i}$ denotes the $i$th coordinate of the vector in $\bbZ^{d}$, and we always have $k_{i} \in \{-1,0,1\}$.
It follows from the assumptions that $k_1=k_4$ and $k_2=k_3$, so that \eqref{diff winding} vanishes.
\qed

The consequence of this proposition is that we effectively lose ergodicity in very large systems, since the dynamics conserves the winding number on simulation time scales. On the other hand, when macroscopic cycles are present, we expect nonzero winding numbers to appear with positive probability in the equilibrium measure. The dynamics always start with the identity permutation which has zero winding number. By the proposition, the path to nonzero winding numbers must cross bottlenecks, i.e, permutations with large jumps which occur with probability less than $\frac1Z \e{-T (\frac L2-1)^{2}}$. A big part of the phase space is not explored by the dynamics.

%We do not expect this to have any effect on the observables discussed in the present paper. 
%Reaching permutations with large jumps requires of order $L$ unfavourable steps which is exponentially unlikely, and such permutations form `bottlenecks' in the state space separating regions with different winding numbers under our dynamics. 
%So although the process is ergodic in theory, in practice the winding number is an observable that is metastable on very long time scales. 
By introducing Monte-Carlo transitions that flip the orientations of cycles (which does not change their probability), it is easy to move between permutations with {\it even} winding numbers. On the other hand, it would be interesting to study the winding numbers of typical permutations at equilibrium, but this seems to be a very difficult task numerically since it requires dynamics that can move between odd an even winding numbers and still sample from the correct distribution.

Due to the metastability of the winding number, the actual mixing time of the Monte-Carlo dynamics is (at least) $\e{c L^{2}}$ with $c>0$. However, since we are only interested in cycle lengths and not in their orientation, we do not expect this to be relevant for the observables discussed in the present paper. In fact, as is shown in Figure \ref{fig mixingtime} and later in Figure \ref{figure:noft}, we observe convergence on time scales of order $|\Lambda |=L^3$. This is still much faster than processes such as card shuffles leading to uniform distributions on permutations (see e.g. \cite{wilson}), which are typically of order $|\Lambda |^3$ with logarithmic corrections. This is due to the fact that for positive temperature $T$ the stationary distribution (\ref{dist}) is not uniform, and jumps in a typical permutation are local of order $1/\sqrt{T}$. Furthermore, the identity permutation we start with is actually the ground state (permutation with highest probability) of that measure.

\section{Effective split-merge process}
In the previous section we presented numerical evidence that the lengths of macroscopic cycles are distributed according to the Poisson-Dirichlet distribution. The goal of the present section is to explain it with a split-merge process, defined below in Section \ref{sec split-merge}. More precisely, the Markov chain Monte-Carlo process of Section \ref{sec MCMC}, when restricted to the cycle structure, becomes an effective split-merge process with the correct rates. We formulate precise conjectures about macroscopic cycles in typical spatial permutations, which we then test numerically.

\begin{figure}[htb]
\begin{center}
\includegraphics[width=6cm]{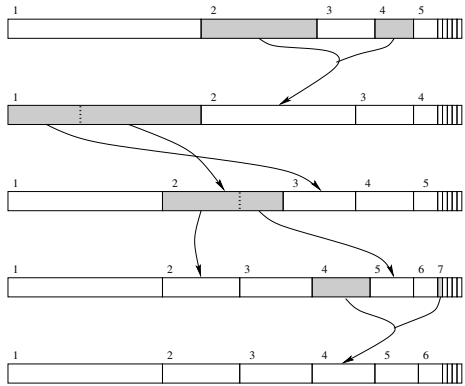}%
\end{center}
\caption{\small Illustration for the split-merge process. The partition undergoes a merge followed by two splits and another merge.}
\label{fig split-merge}
\end{figure}

\subsection{Split-merge process}
\label{sec split-merge}

Recall that a partition $\lambda = (\lambda^{(1)},\lambda^{(2)},\dots)$ of the interval $[0,\nu]$ is a sequence of decreasing positive numbers such that $\sum_{i} \lambda^{(i)} = \nu$. Here, $\nu$ is any positive real number. The split-merge process $(\lambda (t):t\geq 0)$, also called coagulation-fragmentation, is a continuous-time stochastic process on partitions where the $i$th and $j$th components ($i \neq j$) merge with rate
\bea\label{smrates1}
q_{ij} =2\lambda^{(i)}\lambda^{(j)}/\nu^2,
\eea
and the $i$th component is split uniformly into two parts with rate
\bea\label{smrates2}
q_i =(\lambda^{(i)})^2 /\nu^2.
\eea
Note that the rates are in $[0,1]$ and they add up to $1$, so they can be used directly for the following implementation of the process. 
If
$\lambda{(t)} = (\lambda^{(1)}{(t)}, \lambda^{(2)}{(t)},\dots)$ denotes the
partition at time $t$, one chooses the new configuration $\lambda{(t+{\rm Exp}(1))}$ after an exponential waiting time with rate $1$ as follows: %QQ adapted this only slightly
\begin{itemize}
\item Choose a first part of the partition with probability proportional to its
size. That is, the index $i$ is chosen with probability
$\lambda^{(i)}{(t)}/\nu$. This is called ``size-biased sampling''.
\item Choose a second part in the same manner, independently of the first. Let $j$ the corresponding index.
\item If $i \neq j$, merge $\lambda^{(i)}{(t)}$ and $\lambda^{(j)}{(t)}$. That
is, the partition $\lambda{(t+{\rm Exp}(1))}$ contains all parts $\lambda^{(k)}{(t)}$ with
$k \neq i,j$, and a part of size $\lambda^{(i)}{(t)} + \lambda^{(j)}{(t)}$.
%The sequence has been rearranged so that $(\lambda^{(k)}{(t+1)})$ is decreasing.
\item If $i=j$, split $\lambda^{(i)}{(t)}$ uniformly. That is, the partition
$\lambda{(t+{\rm Exp}(1))}$ contains all parts $\lambda^{(k)}{(t)}$ with $k \neq i$, and
two parts $u \lambda^{(i)}{(t)}$ and $(1-u) \lambda^{(i)}{(t)}$, where $u$ is
a uniform random number in $[0,1]$.
\item The sequence is rearranged so that $(\lambda^{(k)}{(t+Exp(1))})$ is decreasing. 
\end{itemize}
The process is illustrated in Fig.\ \ref{fig split-merge}. Additional background can be found in \cite{Ald2,Bert}. Tsilevich showed that the Poisson-Dirichlet distribution is invariant for the split-merge process \cite{Tsi}. It was proved in \cite{DMZZ} that it is the unique invariant measure (see also \cite{Sch}).

\begin{figure}[htb]
\begin{center}
\begin{picture}(0,0)%
\includegraphics{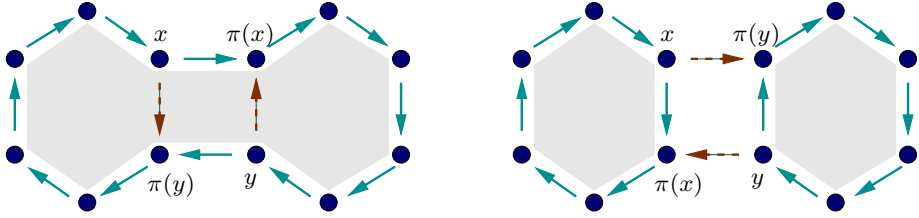}
\end{picture}%
\setlength{\unitlength}{1973sp}%
\begingroup\makeatletter\ifx\SetFigFont\undefined%
\gdef\SetFigFont#1#2#3#4#5{%
  \reset@font\fontsize{#1}{#2pt}%
  \fontfamily{#3}\fontseries{#4}\fontshape{#5}%
  \selectfont}%
\fi\endgroup%
\begin{picture}(11328,2626)(187,-3474)
\put(9451,-3136){\makebox(0,0)[lb]{\smash{{\SetFigFont{9}{10.8}{\rmdefault}{\mddefault}{\updefault}{\color[rgb]{0,0,0}$y$}%
}}}}
\put(1951,-3211){\makebox(0,0)[lb]{\smash{{\SetFigFont{9}{10.8}{\rmdefault}{\mddefault}{\updefault}{\color[rgb]{0,0,0}$\pi(y)$}%
}}}}
\put(2026,-1336){\makebox(0,0)[lb]{\smash{{\SetFigFont{9}{10.8}{\rmdefault}{\mddefault}{\updefault}{\color[rgb]{0,0,0}$x$}%
}}}}
\put(3151,-3136){\makebox(0,0)[lb]{\smash{{\SetFigFont{9}{10.8}{\rmdefault}{\mddefault}{\updefault}{\color[rgb]{0,0,0}$y$}%
}}}}
\put(2926,-1336){\makebox(0,0)[lb]{\smash{{\SetFigFont{9}{10.8}{\rmdefault}{\mddefault}{\updefault}{\color[rgb]{0,0,0}$\pi(x)$}%
}}}}
\put(8251,-3211){\makebox(0,0)[lb]{\smash{{\SetFigFont{9}{10.8}{\rmdefault}{\mddefault}{\updefault}{\color[rgb]{0,0,0}$\pi(x)$}%
}}}}
\put(8326,-1336){\makebox(0,0)[lb]{\smash{{\SetFigFont{9}{10.8}{\rmdefault}{\mddefault}{\updefault}{\color[rgb]{0,0,0}$x$}%
}}}}
\put(9226,-1336){\makebox(0,0)[lb]{\smash{{\SetFigFont{9}{10.8}{\rmdefault}{\mddefault}{\updefault}{\color[rgb]{0,0,0}$\pi(y)$}%
}}}}
\end{picture}%
\caption{\small The transposition $\tau_{xy}$ splits a cycle if $x,y$ belong to the same cycle (left), or it merges two cycles if $x,y$ belong to distinct cycles (right). These are the only two possibilities.}
\label{fig split cycles}
\end{center}
\end{figure}

A key property of our Monte-Carlo process $(\pi_t :t\geq 0)$ described in Section \ref{sec MCMC} is that, at each step, either a cycle is split, or two cycles are merged. This is illustrated in Fig.\ \ref{fig split cycles}. Let us state this precisely.

\begin{proposition}
Let $\pi\in \caS_\Lambda$ and $x,y\in\Lambda$ with $x\neq y$.
\begin{itemize}
	\item If $x,y$ belong to the same cycle in $\pi$, then $x,y$ belong to different cycles in $\pi\circ\tau_{xy}$.
	\item If $x,y$ belong to different cycles in $\pi$, then $x,y$ belong to the same cycle in $\pi\circ\tau_{xy}$.
\end{itemize}
\end{proposition}

See \cite{Kerl} for more details.
It is clear that all cycles of $\pi$ that do not involve $x$ or $y$ are also present in $\pi\circ\tau_{xy}$, and reciprocally. The length of the coalesced cycle is equal to the sum of the lengths of the two original cycles, and similarly for a fragmentation.

\subsection{Effective split-merge process for macroscopic cycles}
\label{sec eff split-merge}

We have seen in the previous section that each Monte-Carlo step results in either splitting a cycle, or merging two cycles. We have also seen in Section \ref{sec nature cycles} that two kinds of cycles are present: The {\it finite} cycles, whose lengths do not diverge in the thermodynamic limit. And the {\it macroscopic} cycles, whose lengths are positive fractions of the volume. A Monte-Carlo step does one of the following:
\begin{itemize}
\item[(a)] Merge two finite cycles.
\item[(b)] Merge a macroscopic cycle and a finite cycle.
\item[(c)] Merge two macroscopic cycles.
\item[(d)] Split a finite cycle (resulting in two finite cycles).
\item[(e)] Splits a macroscopic cycle, resulting in a finite and in a macroscopic cycles.
\item[(f)] Splits a macroscopic cycle, resulting in two macroscopic cycles.
\end{itemize}
One expects each of these options to take place with rates of order
$O(1)$ in the limit $|\Lambda |\to\infty$. The ones that are relevant to the
effective split-merge process are (c) and (f), since their effect is reflected
in a change in $(\lambda^{(1)}(\pi(t)),
\lambda^{(2)}(\pi(t)),\ldots)$, the ordered lengths of \emph{macroscopic}
cycles normalized by $|\Lambda|$. Accordingly, we introduce the rate $R_{ij}$ at
which the $i$th and $j$th largest cycles merge. It depends on the permutation
$\pi$, and, with $i\neq j$, it is given by
\begin{equation}\label{cmerge} %QQ adapted to the convention with 2
R_{ij}(\pi) = 2 \sum_{x \in \gamma^{(i)}} \sum_{y \in \gamma^{(j)}} Q(\pi, \pi \circ \tau_{xy}).
\end{equation}
Notice that $R_{ij}(\pi)$ scales to a constant as the volume diverges, since the sum over $x,y$ is of order $|\Lambda|$, and $Q(\cdot)$ is of order $1/|B(\Lambda)|$, see Eq.\ \eqref{tp}.
The rate $R_{i}$ at which the $i$th largest cycle splits into two macroscopic
cycles involves the cut-off $K$ that distinguishes finite vs macroscopic cycles:
\begin{equation}\label{csplit}
R_{i}(\pi) = %\tfrac12 
\sum_{x \in \gamma^{(i)}} \sum_{k = K}^{L^{(i)}-K} Q( \pi, \pi \circ \tau_{x, \pi^{k}(x)}).
\end{equation}
Here, $\pi^{k}(x)$ is the site at ``distance'' $k$ of $x$ along the cycle $\gamma^{(i)}$.
If we set $K=1$ in the right side, we get the rate at which $\gamma^{(i)}$
splits, irrespective of the sizes of the resulting cycles. The expression above
gives the rate at which $\gamma^{(i)}$ splits in two cycles, each of which has
length greater than $K$. 
As in previous sections we use the cutoff $K = |\Lambda|^{0.6}$.

We expect that, for almost all permutations in equilibrium, the rates $R_{ij}$ and $R_{i}$ are equal to those of the split-merge process modulo a constant time-scale $R$, resulting from the effective rate of macroscopic processes (c) and (f) above. Recall that $\lambda^{(i)}(\pi) = L^{(i)}(\pi)/|\Lambda|$ is a random variable.

\begin{conjecture}\label{conjR}
Let $T$ such that $\nu_\infty(T)>0$. There exists a number $R$ (that depends on $T$
but not on indices) such that for all $i,j$, and all $\varepsilon>0$,
\[
\begin{split}%QQ adapted to factor 2
&\lim_{|\Lambda|\to\infty} \bbP_{\Lambda} \bigl( \bigl| R_{ij} - 2\lambda^{(i)} \lambda^{(j)} R \bigr| > \varepsilon \bigr) = 0, \\
&\lim_{|\Lambda|\to\infty} \bbP_{\Lambda} \bigl( \bigl| R_{i} - (\lambda^{(i)})^{2} R \bigr| > \varepsilon \bigr) = 0.
\end{split}
\]
\end{conjecture}

The time scale of the effective split-merge process is determined by $R$, but the invariant measure is not, so the exact value of $R$ is irrelevant. It is important, however, that it is identical for both the splits and the merges, and for all macroscopic cycles.

\begin{figure}[htb]
\begin{center}

\begin{picture}(0,0)%
\includegraphics{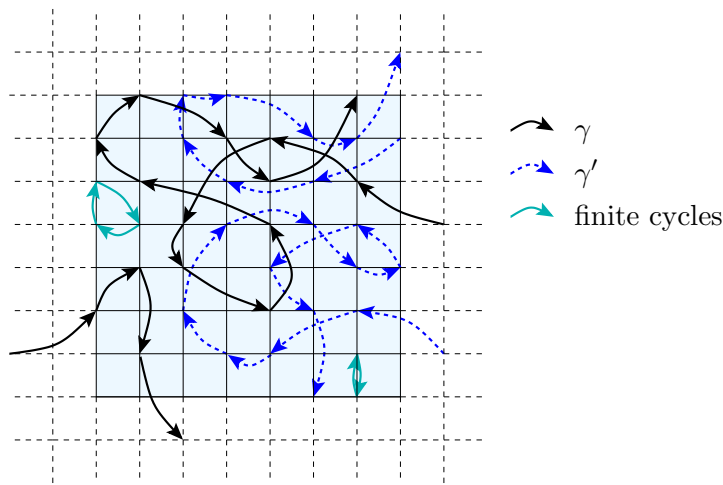}%
\end{picture}%
\setlength{\unitlength}{1776sp}%
\begingroup\makeatletter\ifx\SetFigFont\undefined%
\gdef\SetFigFont#1#2#3#4#5{%
  \reset@font\fontsize{#1}{#2pt}%
  \fontfamily{#3}\fontseries{#4}\fontshape{#5}%
  \selectfont}%
\fi\endgroup%
\begin{picture}(9924,6624)(-32,-5773)
\put(7801,-961){\makebox(0,0)[lb]{\smash{{\SetFigFont{11}{13.2}{\familydefault}{\mddefault}{\updefault}{\color[rgb]{0,0,0}$\gamma$}%
}}}}
\put(7801,-1561){\makebox(0,0)[lb]{\smash{{\SetFigFont{11}{13.2}{\familydefault}{\mddefault}{\updefault}{\color[rgb]{0,0,0}$\gamma'$}%
}}}}
\put(7801,-2161){\makebox(0,0)[lb]{\smash{{\SetFigFont{11}{13.2}{\familydefault}{\mddefault}{\updefault}{\color[rgb]{0,0,0}finite cycles}%
}}}}
\end{picture}%

\caption{\small Schematic drawing of the situation in a mesoscopic box. It
contains finite cycles in light color, and two legs that belong to each of
the macroscopic cycles $\gamma$ and $\gamma'$. The probability that $\gamma$ and
$\gamma'$ merge in the next step is proportional to the number of `contacts'
between them, which in turn is proportional to $L_\gamma L_{\gamma'}$. The
probability that $\gamma$ splits into two macroscopic cycles is proportional to the
self contacts \emph{amongst different legs} of $\gamma$, which is in turn
proportional to $L_\gamma ^2$.}
\label{figure:mesoscopicbox}
\end{center}
\end{figure}

Let us explain the heuristics towards this remarkably simple behavior. Consider
a mesoscopic box $\Lambda'$ whose size is large enough so that boundary effects
are irrelevant, yet small enough so that $\Lambda$ is made up of a large number
of mesoscopic boxes. The restriction of $\pi$ on $\Lambda'$ gives many finite
cycles, and open legs that are parts of macroscopic cycles. See Fig.\
\ref{figure:mesoscopicbox} for a schematic picture. Let us choose a pair of
nearest-neighbors $x,y$ at random, with the condition that $x,y$ belong to
distinct legs. The probability that $\tau_{xy}$ merges $\gamma^{(i)}$ and
$\gamma^{(j)}$ is equal to the probability that $x$ belongs to $\gamma^{(i)}$
and $y$ belongs to $\gamma^{(j)}$, or conversely, which is equal to $2
\lambda^{(i)} \lambda^{(j)} / \nu(T)^{2}$, up to vanishing finite-size effects. %QQ this is already consistent with the new factor 2!
The probability that $\tau_{xy}$
splits $\gamma^{(i)}$ is equal to the probability that both legs belong to
$\gamma^{(i)}$, which is equal to $(\lambda^{(i)} / \nu(T))^{2}$. This
heuristics assumes that macroscopic cycles are spread uniformly in space, so they are
present in all mesoscopic boxes in proportion to their size, and also that the
local configurations do not depend on the situation in other boxes.
Note that short range correlations in the cycle structure affect only the probability that a randomly chosen pair $x,y$ belongs to distinct legs, which is absorbed in the constant $R$ that determines the time scale of the effective split merge process.

In addition, we also conjecture that when a cycle is split, it is split uniformly. 
%The constant $R$ below is the same as the one in Conjecture 1.

\begin{conjecture}
Let $T$ such that $\nu_\infty (T)>0$ and define the CDF for a function of the split length $a\in [0,1]$,
\bea
\label{def theta}
\theta_i^{(a)}(\pi)=\frac{1}{R_i (\pi )}\sum_{x \in \gamma^{(i)}}\sum_{k=K}^{a L^{(i)}}P(\pi,\pi\circ\tau_{x,\pi^k(x)})\ .
\eea
Then
\[
\lim_{|\Lambda|\to\infty} \bbP_{\Lambda} \Bigl( \Bigl| \theta_i^{(a)}(\pi)-a \Bigr| > \varepsilon \Bigr) = 0.
\]
\end{conjecture}

%QQ adapted this paragraph
If these conjectures hold true, the effective process $\bigl( \lambda^{(1)} (\pi (t)),\lambda^{(2)} (\pi (t))\ldots \bigr)_{t\geq 0}$ with stationary initial condition converges in the limit $|\Lambda |\to\infty$ to a split-merge process as defined in \eqref{smrates1} and \eqref{smrates2}, running with total rate $R\,\nu_\infty^2$. Therefore the distribution of cycle lengths has to be invariant with respect to the split-merge process, so it has to be Poisson-Dirichlet. We now check these conjectures numerically.

\begin{figure}[htb]
    
    \begin{center}

        \subfigure{%
            \includegraphics[width=0.4\textwidth]{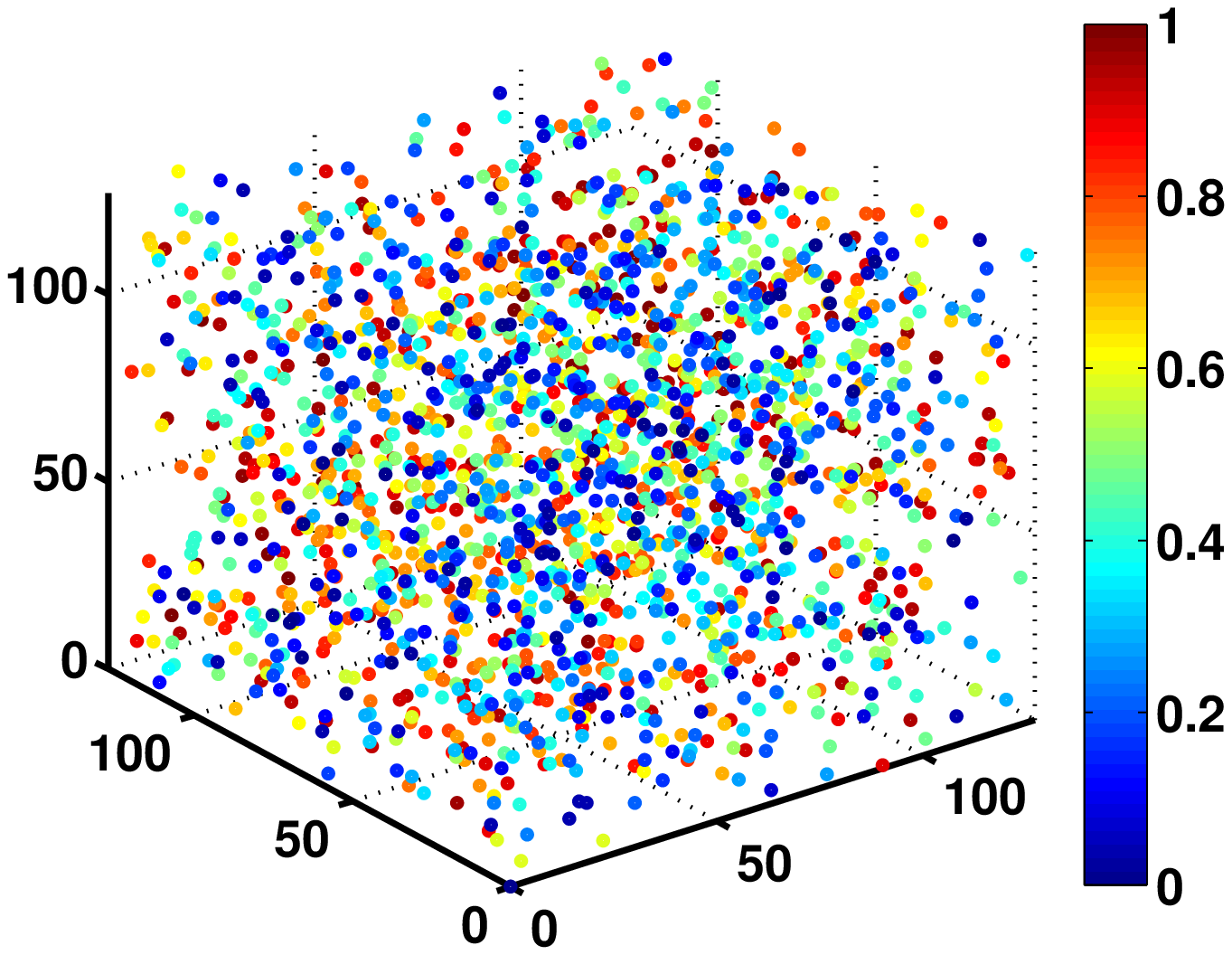}
        }% 
\qquad
        \subfigure{%
           \includegraphics[width=0.4\textwidth]{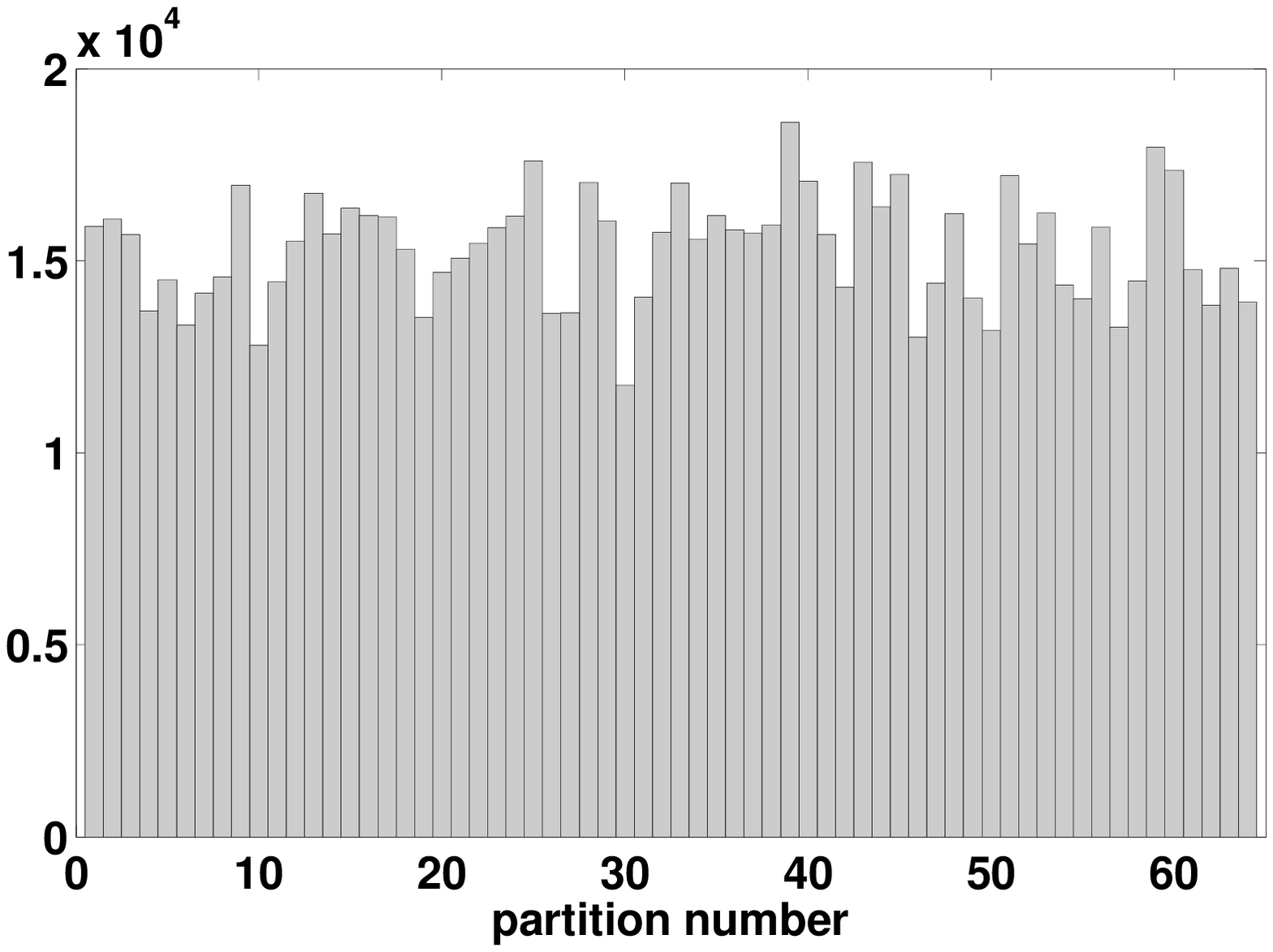}
        } %  ------- End of the first row ----------------------%

    \end{center}
    \caption{\small
A single permutation has been chosen with respect to the equilibrium measure for $| \Lambda | = 128$, $ T=0.8$. The longest cycle $\gamma^{(1)}$ is spread everywhere. (Left) Scatter plot of every 500th site along $\gamma^{(1)}$. The color indicates the distance along $\gamma^{(1)}$, renormalized by the length of $\gamma^{(1)}$, starting from the site closest to the origin. (Right) The box $\Lambda$ of volume $128^{3}$ has been partitioned in 64 subsets of volume $32^{3}$. The histogram depicts the number of sites of $\gamma^{(1)}$ that can be found in each subset. It is essentially constant. 
     }%and a single permutation has been chosen with respect to the equilibrium measure.
\label{fig:mesoscopicboxes}
\end{figure}

\subsection{Numerical data about the effective split-merge process}
\label{sec num split-merge}

In this section we give numerical evidence that the rates for splitting and
merging macroscopic cycles converge to those of a split-merge process, thus confirming
the Poisson-Dirichlet distribution of macroscopic cycles. The heuristics behind this
argument, as described in the previous section, is that macroscopic cycles are distributed uniformly amongst lattice sites on a mesoscopic level, and not are confined to a bounded region. This is illustrated in Fig.~\ref{fig:mesoscopicboxes}.

\begin{figure}[htb]
    \begin{center}
        \subfigure{%
            \includegraphics[width=0.4\textwidth]{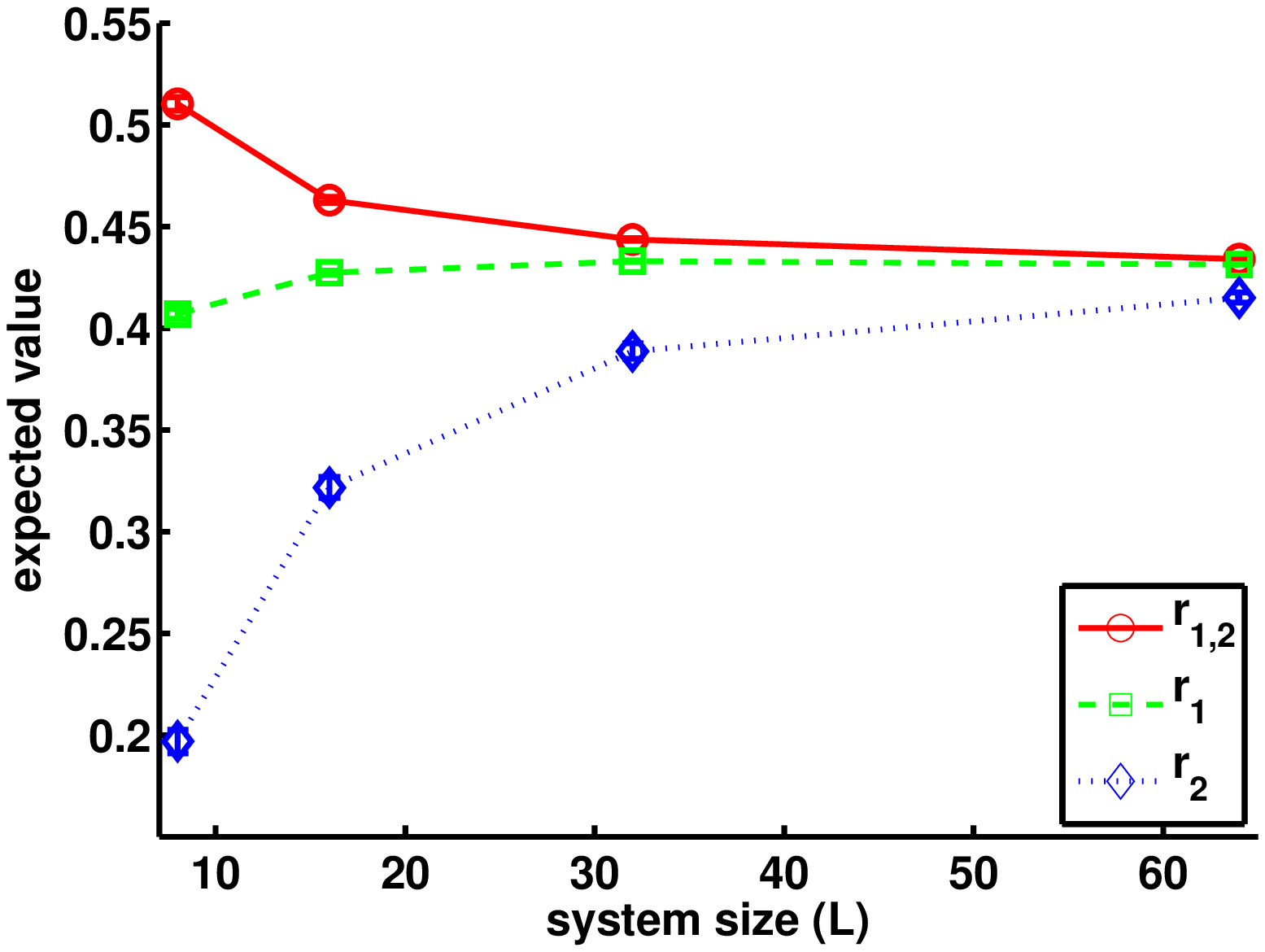}
        }%
	\qquad
        \subfigure{%
           \includegraphics[width=0.4\textwidth]{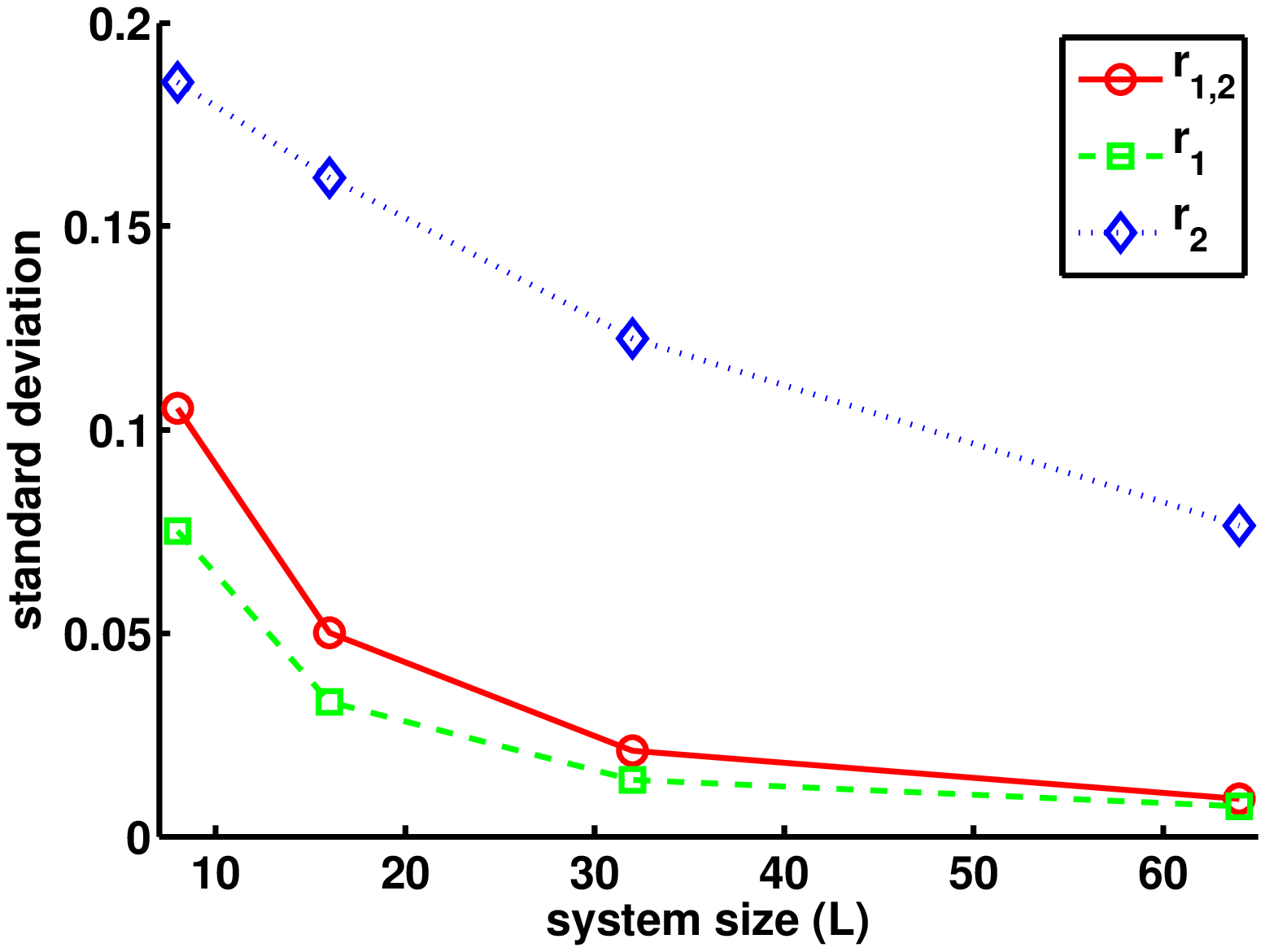}
        } %  ------- End of the first row ----------------------%
    \end{center}
    \caption{\small 
Expectations (left) and standard deviations (right) of the rates defined in \eqref{rates}. $|\Lambda| = 8^{3}, 16^{3}, 32^{3}, 64^{3}$,  $T = 0.8$. The standard error for the mean is within the marker size. Averages were taken over $10^3$ realizations.
     }%
\label{figrates}
\end{figure}

In order to verify Conjecture \ref{conjR},
the rates of merging the two longest cycles, splitting the longest, and
splitting the second longest in the next timestep were calculated using
(\ref{cmerge}) and (\ref{csplit}). We define
\begin{equation}
\label{rates}%QQ adapted to factor 2, Ales should check what he actually used, I think it is anyway the new one
%\begin{split}
r_{ij} (\pi )=\frac{R_{ij} (\pi )}{2\lambda^{(i)} \lambda^{(j)}}\ ,\qquad
r_{i} (\pi )= \frac{R_{i} (\pi )}{(\lambda^{(i)})^2 }.
%\end{split}
\end{equation}
Fig.~\ref{figrates} shows that $r_{ij}$ and $r_{i}$ converge to a constant $R$
as expected, supporting Conjecture 1. Note that in addition to convergence of the mean the variance is decreasing, confirming convergence in probability as stated in the conjecture.

\begin{figure}[htb]
    \begin{center}
        \subfigure[CDF $\theta_1^{(a)}(\pi)$]
        {%
             \includegraphics[width=0.35\textwidth]{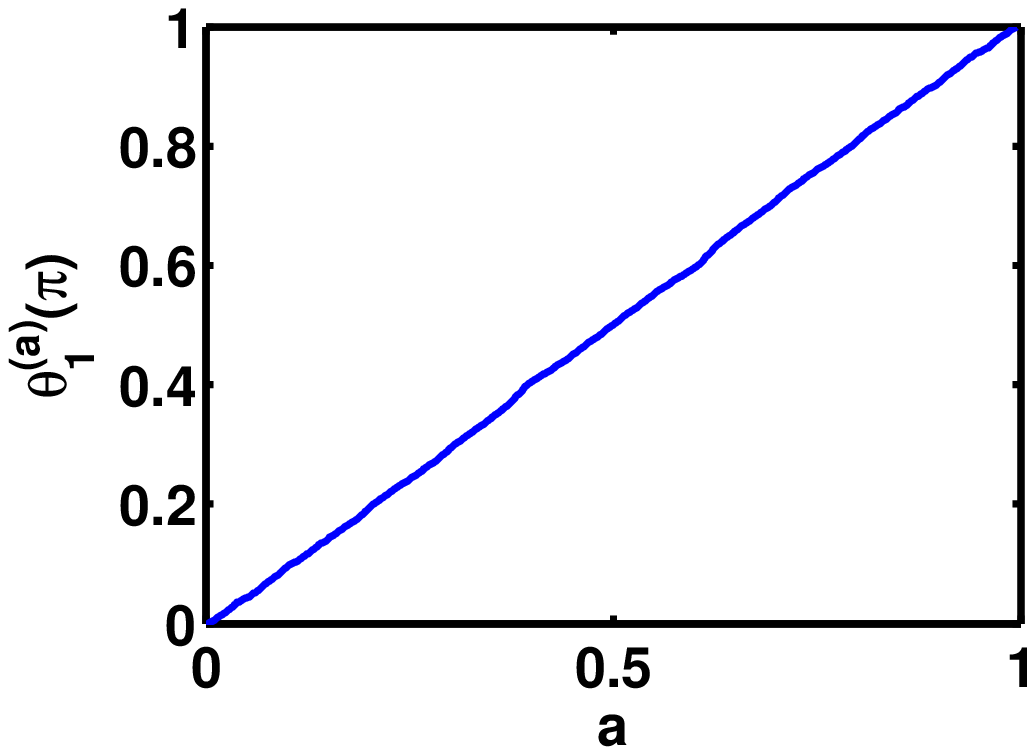}
        }%
\qquad
        \subfigure[Deviations]
        {%
           \includegraphics[width=0.35\textwidth]{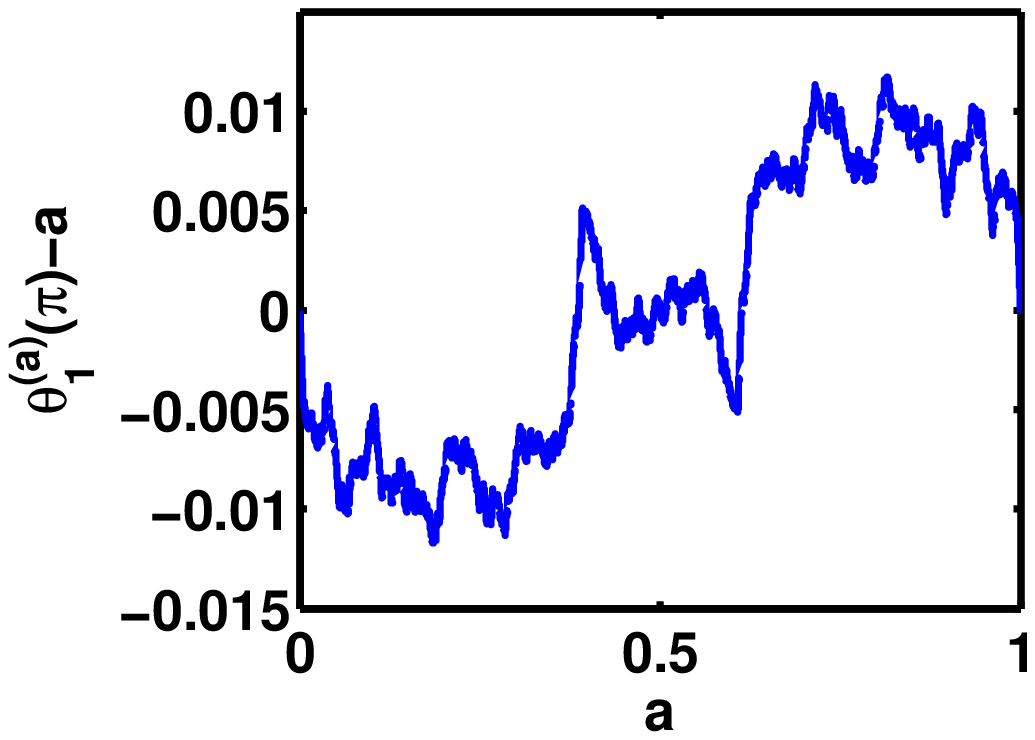}
        } %  ------- End of the first row ----------------------%
    \end{center}
    \caption{\small 
Cycles split uniformly. Plot of $\theta_1^{(a)}(\pi)$ as defined in \eqref{def theta} for a typical permutation $\pi$ under the equilibrium measure with $|\Lambda|=64^{3}$, $T = 0.8$.
     }%
\label{fig uniform split}
\end{figure}

If the cycle is split uniformly, $\theta_i^{(a)}(\pi)$ as defined in Conjecture 2 is the CDF of a uniform random variable on [0,1].
The graph of $\theta_1^{(a)}(\pi)$ is shown in Fig.~\ref{fig uniform split} for a given permutation $\pi$ chosen randomly from the equilibrium measure, which confirms the expected behaviour.

\begin{figure}[htb]
\begin{center}
\includegraphics[width=0.5\textwidth]{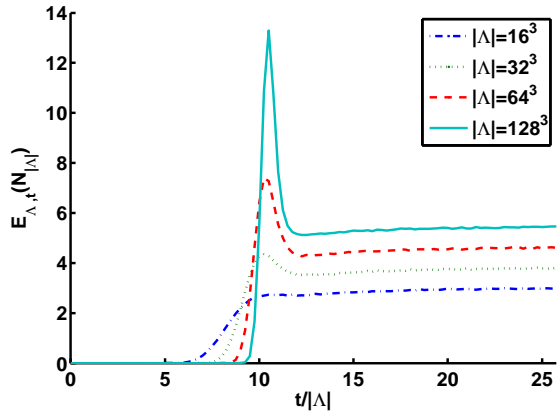} 

\caption{\small Expected value of the number of
cycles longer than $|\Lambda|^{0.6}$, $\bbE_{\Lambda,t}(N_{|\Lambda|})$, as a function of time. The Monte-Carlo chain starts from the
identity permutation and has been recorded for various system sizes. Averages were taken over $10^4$ realizations.}
\label{figure:noft}
\end{center}
\end{figure}

\section{Further prospects}

The emergence of macroscopic cycles seems intriguing and it is worth being studied. Starting the Monte-Carlo Markov chain from the identity permutation, one expects the system to display only finite cycles for some time, before infinite objects built up. Let $\bbE_{\Lambda,t}$ denote the corresponding expectation, and $N_{|\Lambda|}(\pi)$ denote the number of cycles of length larger than $|\Lambda|^{0.6}$. Fig.~\ref{figure:noft} shows $\bbE_{\Lambda,t}(N_{|\Lambda|})$ for $T = 0.8$ and various volumes. There is a peak at the transition to the phase with macroscopic cycles. It is certainly due to the presence of many mesoscopic cycles for a short time, that are going to merge afterwards. It would be interesting to get plots for numbers of cycles larger than $|\Lambda|^{a}$ for $a$ other than $0.6$.

\begin{figure}[htb]
    
    \begin{center}
        \subfigure{%
            %\label{fig:first}
            \includegraphics[width=0.4\textwidth]{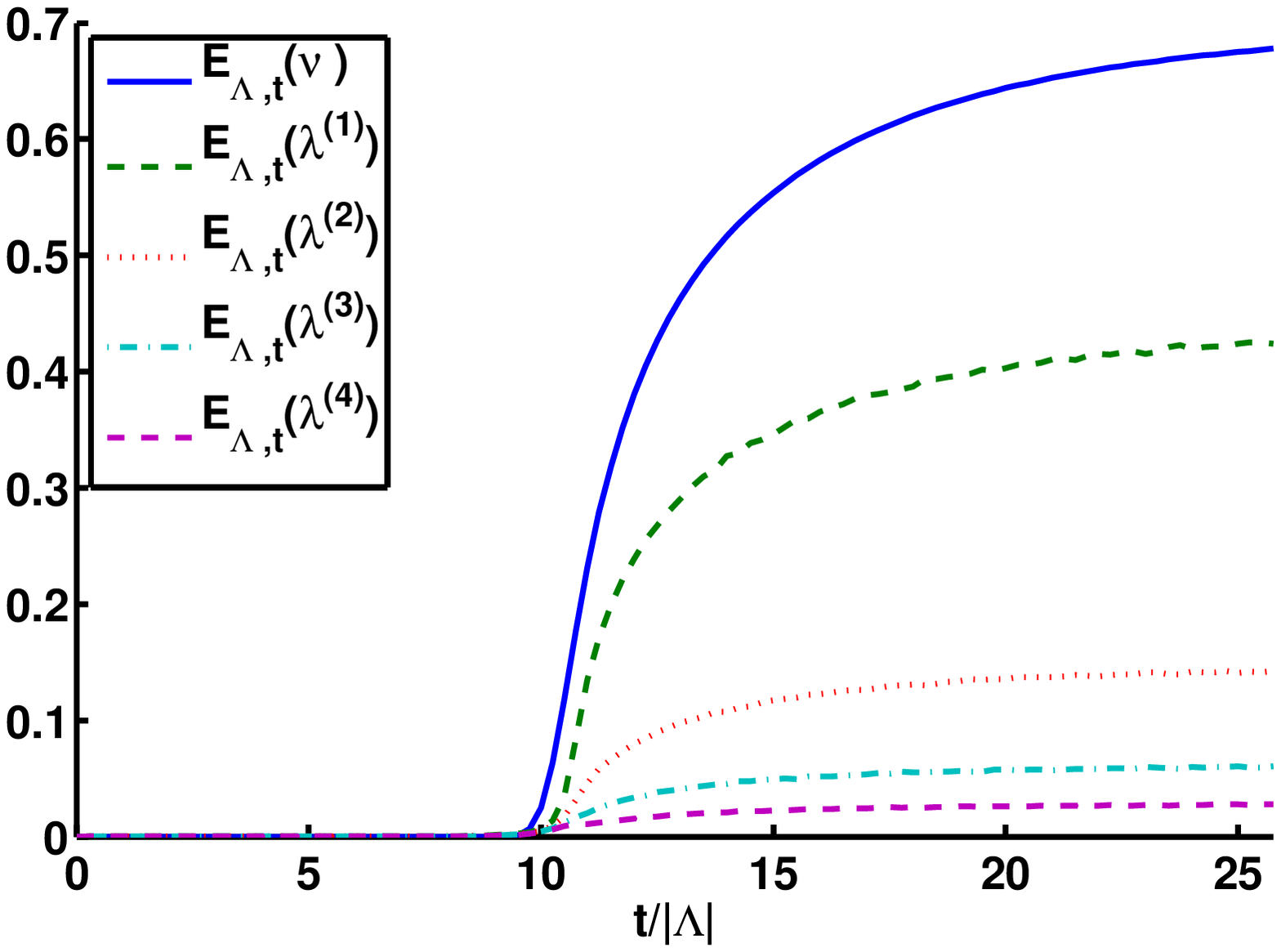}
        }%
\qquad        \subfigure{%
           %\label{fig:second}
           \includegraphics[width=0.4\textwidth]{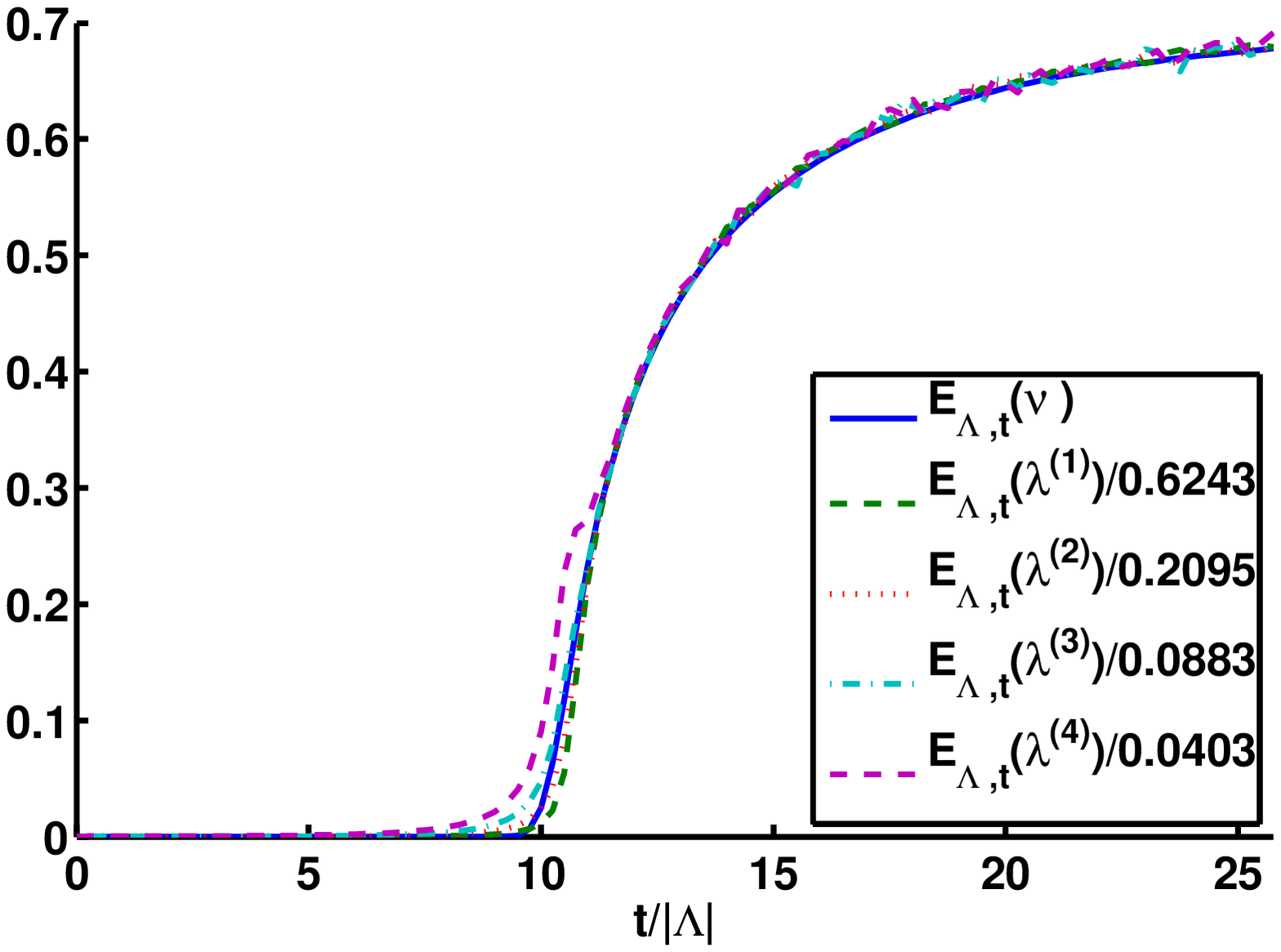}
        }
    \end{center}
\caption{\small The Poisson-Dirichlet distribution occurs already during equilibration. (a) Expected values of the fraction of sites in macroscopic cycles $\bbE_{\Lambda ,t} (\nu )$, and of the lengths of the four largest cycles (divided by the volume). (b) The expectation of the $i$-th longest cycle has been divided by the average length of the $i$-th part in a random partition with Poisson-Dirichlet distribution. It is always close to $\bbE_{\Lambda ,t} (\nu )$. $|\Lambda|=128^3$, $T=0.8$. Averages were taken over $10^4$ realizations.
 }
\label{fig: PD all time}
\end{figure}

While the fraction of sites in large cycles, $\bbE_{\Lambda,t} (\nu )$, varies continuously with time on the scale $t/|\Lambda |$, the steps (c) and (f) of Section \ref{sec eff split-merge} take place at a very high rate (proportional to $|\Lambda |$) once the phase with macroscopic cycles has been reached. One then expects the lengths of macroscopic cycles to split and merge so fast that the Poisson-Dirichlet distribution appears immediately. The numerical results of Fig.\ \ref{fig: PD all time} confirms this. Indeed, the expected lengths of the longest cycles, when divided by the number of sites in macroscopic cycles, is equal to the expected values obtained with respect to the Poisson-Dirichlet distribution, that can be found e.g.\ in \cite{SL}.

Finally, let us comment on the physical dimension, taken here to be $d=3$. It is safe to bet that everything is similar in all dimensions greater than 3. On the other hand, the dimension $d=2$ remains mysterious. There are certainly no macroscopic cycles, as was observed in \cite{GRU}. An open question is whether a phase occurs where a positive fraction of points belong to mesoscopic cycles. This has been ruled out in the ``annealed'' model that involves averaging over point positions \cite{BU1,BU2}. The present lattice model may be closer to a Bose gas with interactions, on the other hand, where a Kosterlitz-Thouless phase transition is expected. The presence of mesoscopic cycles could indeed be related to the slow decay of correlation functions.

\section*{Acknowledgements}

D.U.\ is grateful to N. Berestycki, A. Hammond, and J. Martin for useful discussions. 
A.A.L.\ was funded by the Erasmus Mundus Masters Course CSSM. S.G.\ and D.U.\ acknowledge support by EPSRC, grants no. EP/E501311/1 and EP/G056390/1, respectively.
%\clearpage

\end{document}